\documentclass[11pt,a4paper]{article}
\usepackage{algorithm}
\usepackage[noend]{algpseudocode}
\PassOptionsToPackage{numbers,sort&compress}{natbib}
\usepackage{wavepppreprint}
\usepackage{xurl}
\setcitestyle{numbers,square,comma}
\AtBeginEnvironment{thebibliography}{\interlinepenalty=10000\relax}
\crefname{algorithm}{Algorithm}{Algorithms}

\renewcommand{\floatpagefraction}{0.65}

\renewenvironment{abstract}{%
  \par\vspace{0.75em}%
  \begin{center}%
  \begin{minipage}{0.88\textwidth}%
  \setlength{\parindent}{0pt}%
  \setlength{\parskip}{0pt}%
  {\centering\fontsize{13}{14}\selectfont\bfseries Abstract\par}%
  \vspace{0.65em}%
  \normalfont\fontsize{11}{13.6}\selectfont
  \setlength{\leftskip}{0pt}%
  \setlength{\rightskip}{0pt}%
  \setlength{\parfillskip}{0pt plus 1fil}%
  \ignorespaces
}{%
  \par\end{minipage}\end{center}\vspace{0.95em}%
}

\title{WavePP: High-Throughput Pipeline Parallel LLM Prefill under Prefix Reuse}
\author{%
  \textbf{Aaryam Sharma}\\[2pt]
  {\small Baseten\footnotemark\\[2pt]
  \href{mailto:aaryam.sharma@baseten.co}{aaryam.sharma@baseten.co}}
}
\shorttitle{WavePP}
\hypersetup{
  pdftitle={WavePP: High-Throughput Pipeline Parallel LLM Prefill under Prefix Reuse},
  pdfauthor={Aaryam Sharma},
  pdfkeywords={LLM serving, pipeline parallelism, prefill, prefix caching}
}

\begin{document}
\makebasetentitle
\footnotetext{\url{https://www.baseten.co/}}

\begin{abstract}
Pipeline parallelism can improve prefill throughput by processing multiple request chunks concurrently across different stages of the model. However, keeping the pipeline fully utilized requires efficient scheduling and request preparation. In systems where stages retain and evict cache state independently, a local cache hit does not guarantee that the same prefix can be reused across the pipeline. Here, coordination overhead can impede request admission cadence and thus reduce overall throughput. In this paper, we present WavePP, a prefill runtime built on top of TensorRT-LLM that addresses these challenges by overlapping request admission with pipeline execution. WavePP asynchronously finds a prefix that can be reused across all stages, protects the cached state, and reserves space for the remaining input while earlier requests continue to execute. It subsequently plans the chunk sizes of each request dynamically to maximize pipeline fill. Each stage then completes the local preparation before executing the request. In the same system and pipeline topology, WavePP improves TensorRT-LLM's prefill throughput in 37 of 40 tested settings on GLM~5.2 and MiniMax M2.7. At concurrency 128 with high cache reuse, these changes increase throughput by factors of $2.91$ and $2.02$, respectively. Across 28 Kimi K3 settings, WavePP also has the highest measured throughput in all 18 settings at concurrency eight or higher, compared with tensor/expert-parallel and pipeline-parallel baselines from TRT-LLM, SGLang, and vLLM.
\end{abstract}

\section{Introduction}

Large Language Models (LLMs) have become ubiquitous in our daily lives. As agentic workloads have gained popularity, LLMs increasingly process larger amounts of text in the conversation, documents, and outputs from tools. Even if only a few tokens are generated, serving such requests can be computationally expensive \citep{srivatsa2025preble}. Therefore, improving inference performance has become a key focus of recent serving systems \citep{kwon2023efficient,sglang,sarathiserve}.

LLM inference is often split into a prefill phase and a decode phase. The prefill phase processes the initial input tokens and fills the cache with the necessary state for generation, while the decode phase autoregressively generates new output tokens, continuously updating the same cache. In many enterprise systems, this work is often split between separate worker types using disaggregated serving. This is especially effective when multiple long context requests need to be processed so that the prefill computation remains decoupled from the decode work. 

In addition to increasing input sizes, however, LLMs themselves have also substantially grown in size. The largest frontier models often do not fit on a single GPU. Including space for caches, models are often split over even more GPUs. Several parallelism strategies exist to achieve this, such as tensor parallelism (TP) which divides individual layers across GPUs, pipeline parallelism (PP) splits the model into contiguous groups of layers, and context parallelism (CP) divides work along the request input length. Each strategy has different costs in terms of communication, but also the concurrent amount of work that is done \citep{shoeybi_megatron-lm:_2020,GPipe,liu2024ringattention}. In particular, each parallelism strategy not only has specific advantages for different workloads, but also for different inference phases.

PP has often been used for throughput efficiency in both training and inference. Since several microbatches are passed through successive groups of layers in parallel, it allows several parts of the model to execute concurrently, also increasing the number of concurrent inputs processed \citep{GPipe,narayanan_pipedream:_2019,revisitingpp}. Many training systems hence also support PP for general neural networks. For inference, many popular libraries support pipeline parallelism for prefill \citep{sglang_pipeline,vllm_architecture} and several works optimize scheduling when prefill and decode share a pipeline. These optimizations include improving scheduling to reduce bubbles between prefill and decode batches \citep{sarathiserve}, dynamically chunking requests to reduce imbalance between pipeline stages \citep{revisitingpp}, and overlapping host-side work with GPU work efficiently for prefill or decode \citep{guo_gllm:_2025,he_sipipe:_2025}.

Due to the autoregressive nature of most modern LLMs, the cached states of a shared prefix can be reused across requests which greatly reduces computation in online serving, agentic workflows, and batch inference with common shared prompts \citep{pan_marconi:_2025,kwon2023efficient,sglang,zhang_jenga:_2025,liu_lmcache:_2025}. For PP, each stage typically contains a local cache for its layers, making request preparation a distributed operation. This coordination can be simplified with a shared cache view, or by replicated schedulers operating on compatible logical states in practical inference systems. Instead, we consider an alternative system, where KV blocks and recurrent-state snapshots may be retained, offloaded, and evicted independently, both within a stage and also across pipeline stages. This enables each stage to conduct host side cache management independently, while ranks agree on reuse and follow a common execution schedule. Each cache manager can respond to local memory pressure and prepare requests around local execution. When pipeline stages are not balanced, one stage can prepare a request while another is still computing.

However, this means that in our setting, a local cache hit does not guarantee that this point can be reused globally. The ranks hence must agree on a common reuse boundary, protect the agreed upon state, and reserve sufficient capacity for the uncached input suffix. Performing this work on the main execution path can delay the next forward pass, meaning that our runtime must establish these guarantees sufficiently early to maintain pipeline occupancy.

We thus present WavePP, a runtime design that establishes this admission agreement efficiently while earlier requests continue through the pipeline. Each stage protects the reusable state and reserves space for the rest of the input before the request is scheduled. The stages then finish preparing their local caches independently, allowing stage 0 to begin without waiting for every stage to finish. The executor sees a sequence of \emph{waves}, each containing one or more request chunks, hence leading to our name.

Our contributions are:

\begin{itemize}
  \item We design WavePP, an asynchronous prefill runtime that overlaps request admission with pipeline execution while allowing stages to manage their caches independently.
  \item We develop a protocol that agrees on reusable state and reserves space for the remaining input before allocating the uncached suffix's blocks. This lets stages prepare requests independently and delays eviction until the reserved space is needed.
  \item We show that faster admission, more waves in flight, and adaptive wave sizing together improve throughput on the same kernels. Within TensorRT-LLM PP4, these changes improve throughput in 37 of 40 GLM~5.2 and MiniMax M2.7 settings, with $2.91\times$ and $2.02\times$ throughput for short-suffix requests under cache pressure at concurrency 128.
  \item We compare WavePP with TP and PP baselines across cold, cached, and mixed Kimi K3 workloads. It has the highest measured throughput in all 18 settings at concurrency eight or higher.
\end{itemize}

\section{Related Works}

\subsection{Pipeline Parallelism}

Pipeline parallelism has been extensively studied for training. GPipe partitions a model across devices, and accumulates gradients before each training update \citep{GPipe}. PipeDream also overlaps work from different minibatches, using weight versions to support pipelined forward and backward execution \citep{narayanan_pipedream:_2019}. More recently, LLMs also have several frameworks for efficient PP training \citep{narayanan_efficient_2021,qi_zero_2023}. TeraPipe further uses a dynamic programming-based algorithm to optimize the pipeline execution in training \citep{terapipe}.

While there is no backward pass in inference, pipeline parallelism can still be beneficial for throughput. Sarathi-Serve combines chunked prefill with stall-free batching to schedule prefill alongside concurrent decode passes and reduce variation between microbatches \citep{sarathiserve}. gLLM separately regulates prefill and decode token counts, where its prefill budget accounts for both pending work and KV-cache availability \citep{guo_gllm:_2025}. Revisiting Pipeline Parallelism dynamically adjusts chunk sizes and delays selected decode requests to reduce imbalance \citep{revisitingpp}. WavePP also adjusts the wave token budget, while focusing on preparing requests early enough to keep the executor supplied with work.

Several works focus on improving execution around each forward pass, for both decode and prefill. gLLM sends metadata ahead of activations so workers can prepare inputs during computation \citep{guo_gllm:_2025}. TD-Pipe temporally separates prefill and decode for offline inference, balances decode batches, and separates centralized scheduling from distributed execution \citep{zhang_td-pipe:_2025}. SiPipe moves sampling to CPUs and overlaps input preparation with GPU work \citep{he_sipipe:_2025}. VPP uses a virtual-stage arrangement for long-context prefill, reducing imbalance while keeping chunk sizes fixed \citep{shi_vpp:_2026}.

Many practical inference libraries also support PP. In vLLM's V1 PP runtime, an engine-core scheduler and cache manager coordinate the distributed workers \citep{vllm_architecture}. SGLang uses replicated schedulers over an ordered request stream \citep{sglang_repository}. Both support overlapping pipeline execution, including successive prefill chunks of the same request in flight \citep{sglang_repository,vllm_repository}. WavePP focuses on distributed admission where it agrees on reusable state, protects that state, and reserves suffix capacity outside the executor loop, while each stage manages its local cache, reducing the effect of admission on the execution cadence.

Another common direction of improvements to PP is in the allocation of layers across pipeline stages. Some approaches enable live redistribution of layers in real time, including migration of cache across stages such as in DynaPipe and PipeLive \citep{xu_dynapipe:_2025,bai_pipelive:_2026}. Other approaches such as Helix optimize static model placement and request scheduling across heterogeneous devices as an optimization problem \citep{mei_helix:_2025}. WavePP studies admission within a fixed pipeline configuration.

\subsection{Cache Reuse and Serving Runtimes}

Prompt caching and reuse have been established as powerful accelerators across various works \citep{gim_prompt_2024}. Preble jointly considers prefix reuse and computation load when assigning requests across workers \citep{srivatsa2025preble}. Most serving runtimes support cross-request cache reuse. PagedAttention in vLLM manages KV state in blocks \citep{kwon2023efficient}, while RadixAttention in SGLang organizes reusable prefixes in a radix tree \citep{sglang}. TensorRT-LLM also builds a radix tree in order to coordinate prefix reuse efficiently \citep{tensorrt_kv_cache}. WavePP is implemented on top of TensorRT-LLM, adopting its caching strategy.

These caches are often also distributed across many layers of memory, including on-device, on host, disk, and remote storage. Tools such as Mooncake and LMCache enable this distributed storage, while also supporting cache transfers \citep{mooncake,liu_lmcache:_2025}. Modern LLMs also combine attention types with different cache mechanisms, including recurrent layers whose prefix caches retain selected state snapshots. Marconi selects cache entries using their expected reuse and compute savings relative to memory cost \citep{pan_marconi:_2025}. Jenga supports heterogeneous cache allocation and caching policies tailored to the dependencies of each layer \citep{zhang_jenga:_2025}. Hybrid cache management is also supported in vLLM, SGLang, and TRT-LLM \citep{vllm_hybrid_cache,sglang_unified_radix,tensorrt_llm}.

Enterprise systems also employ disaggregated serving, separating prefill and decode work across different worker types to reduce interference and meet different latency objectives \citep{distserve,splitwise}. Mooncake builds a disaggregated serving architecture around distributed KV storage and reuse \citep{mooncake}. Baseten's discussion of the inference efficiency frontier describes how dedicated workers allow the two phases to use different configurations and scale with the workload \citep{baseten_efficient_frontier}. Disaggregation is also implemented in TensorRT-LLM, vLLM, and SGLang, which provide mechanisms for transferring cache state between prefill and decode workers \citep{tensorrt_disaggregated,vllm_disaggregated,sglang_disaggregated}. NVIDIA Dynamo coordinates disaggregated worker pools and KV-aware routing across these inference backends \citep{nvidia_dynamo}. WavePP thus focuses on a prefill-only worker.

The terms lease and escrow have a longer history in distributed systems. Gray and Cheriton's leases give time-limited rights over cached data \citep{gray_leases:_1989}. WavePP uses the term for a local reference that protects a cached prefix, without time-based expiry. O'Neil's escrow method reserves quantities for concurrent transactions \citep{oneil_escrow_1986}. Our capacity escrow similarly reserves a shared resource before use, but does not implement database transaction or recovery semantics.

\subsection{Other Parallelism Strategies}

PP can be combined with other ways of dividing model execution. Megatron-LM develops tensor parallelism, while DeepSpeed Ulysses, Ring Attention, and USP distribute long-sequence attention work across devices \citep{shoeybi_megatron-lm:_2020,jacobs_deepspeed_2023,liu2024ringattention,fang_usp:_2024}. Expert parallelism distributes the experts of mixture-of-experts models \citep{lepikhin2021gshard}. Helix Parallelism combines KV sharding during attention with tensor parallelism for dense feed-forward layers, or tensor and expert parallelism for MoE layers \citep{bhatia_helix_2025}. These strategies change how work and state are distributed within layers and can complement a pipeline across layer groups. We compare PP against tensor parallelism in this paper.

\section{Preliminaries}
\label{sec:preliminaries}

\subsection{Transformer Architectures}

Autoregressive attention typically produces various intermediate representations that are reusable across subsequent generation steps of a request. In conventional multi-head attention (MHA), each layer retains the keys (K) and values (V) associated with its KV heads, which allows subsequent tokens to attend to the previous tokens without recomputing them. Attention mechanisms such as grouped-query and multi-query attention reduce KV storage by sharing keys and values between query heads \citep{ainslie2023gqa,shazeer2019mqa}. Multi-head latent attention (MLA) instead stores a lower-dimensional latent representation from which the attention content keys and values can be reconstructed \citep{deepseekai2024deepseekv2}. The storage mechanisms for these representations are collectively referred to as the \emph{KV Cache}.

Typical attention mechanisms such as MHA or MLA are $O(n)$ in computation for the $n^{\text{th}}$ token in the sequence, and have a growing KV cache that scales with the sequence length. Linear attention architectures replace this growing KV cache with a recurrent state whose dimensions are independent of sequence length. A broad family of linear attention mechanisms can be written as
\begin{equation}
    S_t = A_t S_{t-1} + U_t, \qquad o_t = f(q_t, S_t),
\end{equation}
where $S_t$ captures the prefix $x_{1:t}$. This recurrent expression is a central property of linear attention \citep{katharopoulos2020transformers,yang_gated_2025}.

Kimi Delta Attention (KDA), used in Kimi K3, is one such architecture \citep{team_kimi_2025,team_kimi_2026}. KDA builds on Gated DeltaNet, combining delta-rule state updates with finer-grained gating \citep{yang_gated_2025,team_kimi_2025}. A simplified single-head KDA update can be written as
\begin{equation}
    S_t = \left(I-\beta_t k_t k_t^\top\right)\operatorname{Diag}(\alpha_t)S_{t-1} + \beta_t k_t v_t^\top,
\end{equation}
followed by
\begin{equation}
    \widetilde{o}_t = S_t^\top q_t.
\end{equation}

\subsection{Reuse}

Two requests with the same common prefix can share the computed cache for that prefix. Cache reuse can speed up request prefill by several times. For example, a 100k context request with 90\% cached only requires processing the last 10k token suffix. Hence, most popular inference serving systems store the model's cache for reuse across requests \citep{sglang,tensorrt_kv_cache,vllm_prefix_caching}.

While the KV cache saves computation by reusing previous intermediate results, it can become expensive on memory to store it during long-context serving. Instead of storing a single contiguous tensor for each request's KV cache, modern inference engines typically allocate the cache in smaller, fixed-size pages, as in PagedAttention \citep{kwon2023efficient}, commonly also referred to as KV-cache blocks. A request containing $n$ cached tokens and a block size of $B$ tokens occupies approximately $N_{\mathrm{blocks}}(n) = \left\lceil \frac{n}{B} \right\rceil$ logical blocks, whose physical storage need not be contiguous. Paging allows blocks to be allocated on demand, reduces memory fragmentation, and enables block-level sharing between requests.

\emph{Prefix caching} is the mechanism that allows requests to identify and reuse these blocks across requests. SGLang's RadixAttention, for example, organizes shared token prefixes using a radix tree \citep{sglang}. Other implementations, such as in vLLM, identify cached blocks using hashes derived from the tokens within a block together with the preceding prefix \citep{vllm_prefix_caching}. An implementation may therefore index token prefixes directly, index block-sized token sequences, or use a chained block hash.

While GPU memory is the fastest location for cached blocks, inference systems often extend this KV storage into CPU DRAM, local SSDs, or even remote memory. Mooncake, for example, uses CPU DRAM and SSD capacity as part of a distributed KV-cache hierarchy \citep{mooncake}, while LMCache provides an external cache layer capable of moving KV state among GPU memory, CPU memory, disk, and remote storage \citep{liu_lmcache:_2025}. Such hierarchical caching increases the cache capacity, and transporting the cache can be much faster than recomputing it. The logical existence of a cached prefix, hence, differs from its physical residency in the system at any given moment.

Linear attention changes prefix reuse fundamentally. To resume a full-attention layer after token $r$, the runtime needs all KV cache blocks corresponding to the previous tokens. On the other hand, to resume a recurrent layer, the recurrent state after exactly that prefix, $S_r$, is required. In general, a later state $S_{r'}$ for $r'>r$ cannot be rewound to recover $S_r$ \citep{pan_marconi:_2025}.

While a single recurrent state is constant with respect to sequence length, an individual state snapshot can be significantly large. Ignoring smaller auxiliary states, the storage required by one checkpoint is approximately
\begin{equation}
    M_{\mathrm{snapshot}} \propto b \sum_{\ell \in \mathcal{L}_{\mathrm{linear}}} H_\ell d_{k,\ell}d_{v,\ell},
\end{equation}
where $b$ is the number of bytes per state element, $H_\ell$ is the number of state heads, and $d_{k,\ell}$ and $d_{v,\ell}$ are the K, V dimensions. Hence, materializing a snapshot at every token or at frequent intervals such as KV block boundaries becomes infeasible for long context serving. Practical recurrent-state caches therefore store snapshots only at selected boundaries, such as sparse periodic intervals or request boundaries \citep{pan_marconi:_2025}.

Models such as Kimi K3 contain MLA layers as well as linear attention layers. These heterogeneous state types can be managed in a single unified cache, or they can be managed independently. The former design allows a consistent view across caches, while the latter allows each cache to use retention and storage policies appropriate for its state type. Current vLLM coordinates multiple cache groups, including recurrent or Mamba and full-attention groups, through a common cache coordinator and shared block pool \citep{vllm_hybrid_cache}.

\begin{figure}[!htbp]
    \centering
    \includegraphics[width=\linewidth]{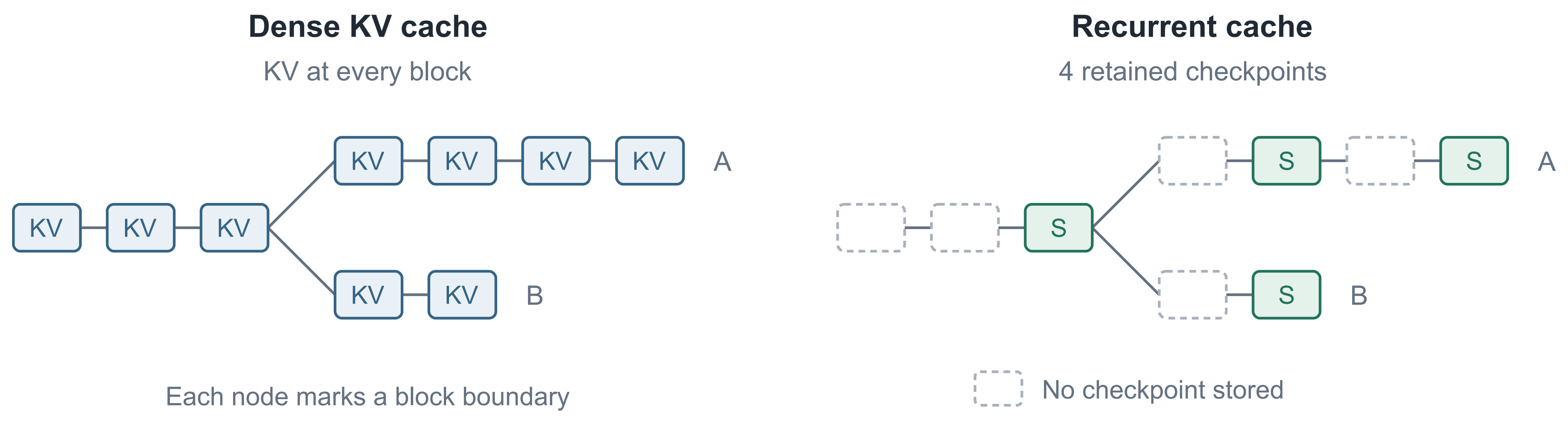}
    \caption{Dense KV storage and sparse recurrent checkpoints on the same prefix tree. The recurrent cache retains four snapshots ($S$): one at the fork, two on branch A, and one at the end of branch B. Dashed nodes mark boundaries without a stored snapshot.}
    \label{fig:reuse-cache-types}
\end{figure}

When these caches evolve independently, the reusable prefix boundaries may not be consistent, especially because the linear attention checkpoints are sparse. Thus, the reusable prefix boundary for a request must belong to the intersection of the reusable boundaries across the two caches. Figure~\ref{fig:reuse-cache-types} illustrates this sparsity on two identical prefix trees.

\subsection{Disaggregated Serving}

During the \textbf{prefill} phase of inference, the uncached input tokens are processed and the model materializes the state needed to continue generation. During the \textbf{decode} phase, the model autoregressively processes the newest token while reading and updating the state created by the preceding prefix until a stopping condition is met, thereby generating the output.

Disaggregation allows these two phases to use different parallelism strategies and worker allocations, since their computation and memory requirements differ. The state produced by prefill is transferred to a decode worker before generation continues \citep{distserve,splitwise,baseten_efficient_frontier}.

Our focus is on prefill workers in a dedicated disaggregated serving system that we run using pipeline parallelism. Our primary metrics are token throughput and prefill completion time (latency), as these are key factors in service level agreements (SLAs).

\subsection{Parallelism Strategies}

\begin{figure}[!htbp]
    \centering
    \includegraphics[width=\linewidth]{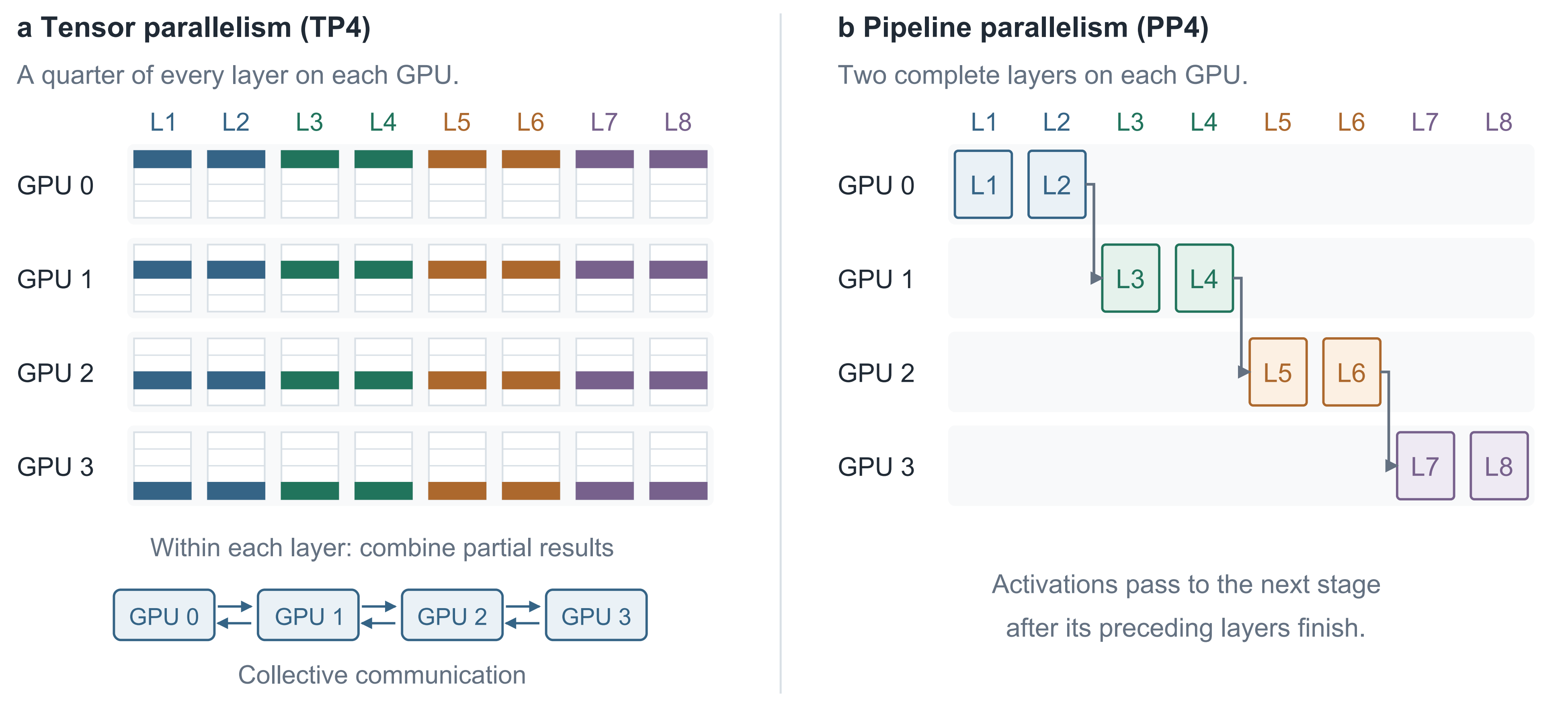}
    \caption{Two ways to divide an eight-layer model across four GPUs. TP splits each layer among the devices, combining partial results through collectives. PP assigns complete layer groups to successive stages and passes activations between them.}
    \label{fig:parallelism}
\end{figure}

Suppose that the model has $L$ layers and we execute on $P$ GPUs. Figure~\ref{fig:parallelism} illustrates how TP and PP divide the model across them.

Under \textbf{tensor parallelism} (TP), each layer is divided among the $P$ ranks. Each rank processes the same token batch and exchanges partial layer results through collective communication. This can lead to fast execution of individual layers, but requires collective communication between devices for each sharded layer, which also adds overhead \citep{shoeybi_megatron-lm:_2020}.

Under \textbf{pipeline parallelism} (PP), the layers of the model are partitioned among $P$ ordered stages. Stage $s$ owns a contiguous fixed subset $L_s$ of the model. Each stage processes its own assigned layers and on completion of the forward pass, it sends its output activations to stage $s+1$. Different microbatches of requests can occupy different stages concurrently. Each PP stage owns the cache state belonging to its layers.

Since each stage must wait for the previous stage to finish processing a microbatch before it can start, the overall throughput is limited by the slowest stage. This can lead to underutilization of resources if the workload is not balanced across stages. Online serving further complicates the problem because the cost of processing a microbatch can vary significantly due to factors such as cache hits, prompt lengths, and chunk tails. Therefore, a PP scheduler must expose enough independent work to fill the stages while keeping the work in successive microbatches balanced.

PP can also enable more cache space when TP replicates caches across ranks. However, PP requires consistency in cache reuse across stages. The host-side and scheduler work increases as more requests are in-flight. Moreover, the system should be designed carefully to prevent gaps across the pipeline due to slow admission rates. Thus, despite the numerous advantages of PP, achieving consistent and efficient execution is challenging, especially in an online inference setting where requests arrive asynchronously and have varying lengths.

\section{Design of WavePP}
\label{sec:design}

In this section we present the design of WavePP. We first describe the steps that must be taken before a request can execute, then explain how this work is overlapped with earlier requests. Finally, we describe how the scheduler divides the admitted work to keep the pipeline occupied.

\paragraph{Waves and lanes}
After admission, a request's uncached suffix is divided into contiguous chunks for processing as in chunked prefill. We call a microbatch of one or more request chunks that move through the pipeline together a \emph{wave}. Each request keeps a logical place, or \emph{lane}, across consecutive waves until all its chunks are scheduled.

\subsection{A Minimum Implementation}
\label{sec:design-minimum}

A straightforward implementation in our setting can establish consistent prefix reuse and allocate space for the remaining context through the following steps:
\begin{enumerate}
  \item Receive and preprocess the request (such as tokenization), then transport it to every rank so each stage can prepare its local state.
  \item Establish a reuse agreement across all cache types and ranks.
  \item Pin the local cache for reuse, including recurrent snapshots, and allocate sufficient memory for the uncached suffix for the request.
  \item Build the request execution metadata including materialization of the request's uncached suffix blocks in the cache and references to cache memory.
  \item Chunk and execute the requests using chunked prefill. Each stage receives activations from its predecessor, which are used for the current layer's computation and then transferred to the successor.
  \item Publish the completed cache state for reuse for future requests, and transfer the cache to a decode worker, or continue decoding locally.
\end{enumerate}

In this implementation, every rank receives the request tokens to update its local cache tree. The ranks need a common reuse boundary so that each stage has the context required for the selected suffix. A simple implementation can serialize admission to keep requests from claiming or evicting each other's cache. To admit several requests concurrently, it must protect both reusable state and suffix capacity. If requests pin most of the cache and then wait for more space, none may be able to finish without releasing state or being preempted. Each stage also needs the activations from the previous stage before computing, and cache publication makes the completed work reusable by future requests.

Furthermore, many of these steps require mutual exclusion, since concurrent modifications to the cache could lead to inconsistencies or race conditions. In a simple two-walk implementation, each rank first traverses its radix trees and communicates its maximum reuse endpoint. Once a common endpoint is established, the ranks traverse their trees again to validate and pin the cache for reuse, protecting those branches from eviction. The tree mutex serializes changes such as eviction and pinning, and prevents two requests from claiming the same free block or eviction victim. This protects the allocation itself, but does not ensure that enough capacity remains for every admitted request to finish.

However, under pressure of several long context requests, doing such a traversal becomes costly, and since ranks may diverge, a second walk post-communication to rectify the current cache lease adds even more cost. Under first-come, first-served scheduling this can be even more cumbersome when the request must requeue to access the radix tree mutex, or other requests must block on the first one completing its communications.

Another consideration is that we want waves of newer requests to schedule while previous waves are computing. Alongside decoupling reuse estimation from the main execution loop, we allow each rank to materialize its cache locally. This lets stage 0 begin once it is ready while later ranks still prepare the request.

\subsection{Decoupling Admission}

We first decouple admission from the main execution of the model. Each rank has two main threads. The \emph{WAVE} thread transports requests and handles cache reuse agreement and capacity. The \emph{LOOP} thread runs the executor, including the materialization of request cache blocks. Materialization uses the local allocator to claim blocks and build the sequence metadata, including radix-tree attachments. Rank 0 is the leader rank, and does the chunk planning on the executor thread.

We also split communication across two planes. The \emph{data plane} carries wave schedules and activations across stages to support the model forward, while the \emph{rounds plane} handles requests and admission results using a rank 0 broadcast followed by an all-gather. Communications rounds contain multiple requests at once to amortize overhead. Admission rounds run more frequently while waiting for another rank's response, and repeat installation records only for requests still being admitted. Though these collectives may block the admission thread (WAVE), the LOOP thread can continue executing earlier admitted work. Data-plane sends and receives are asynchronous, allowing transfers to overlap work that does not depend on their completion. Each stage still waits for the inputs needed by its next forward pass.

Rank 0 receives the request, and announces it once it enqueues. The other ranks register this and prepare the request locally, which allows hashing and reuse negotiation to begin while the executors continue previous work. The cache agreement is separate from physically materializing the state in the tree. Before we schedule, every rank must only promise that it has pinned the agreed upon cache (lease), and claimed sufficient space for the uncached suffix (escrow), i.e., it does not need to add the blocks to the respective cache trees. Later, each rank completes its own materialization before the wave is executed. This means that stage 0 can execute the request once locally ready, without waiting for future ranks to complete installation. A later rank that is not ready when the wave arrives can still introduce a local only gap. On follower ranks, kernel launches can block while holding the Python interpreter lock. We therefore wait for a receive event with that lock released before enqueueing the next forward. The executor also yields when there are requests awaiting admission without runnable work.

\begin{figure}[!htbp]
    \centering
    \includegraphics[width=\linewidth]{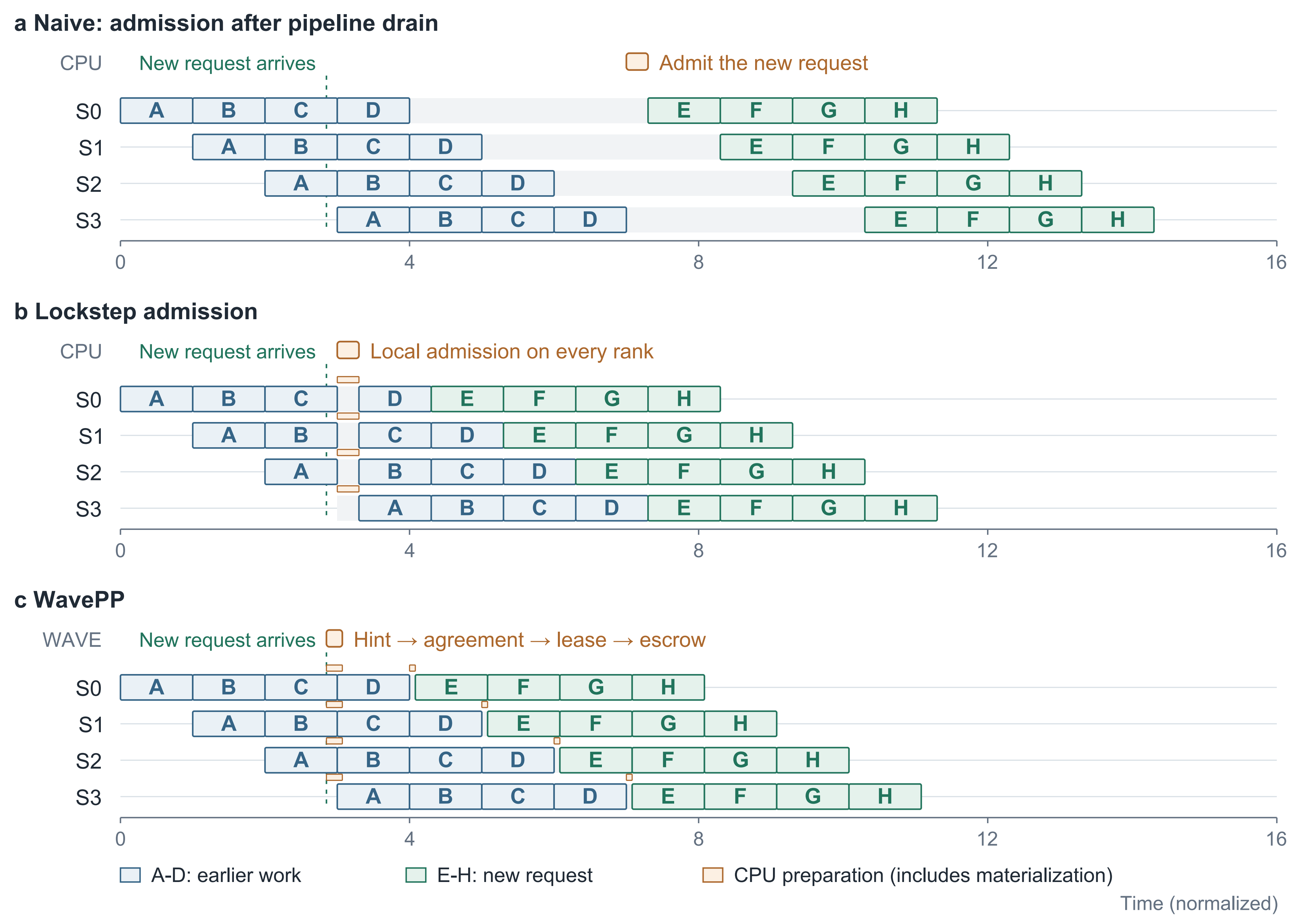}
    \caption{A new request arrives while stage 0 executes C. The naive policy waits for the pipeline to drain. Lockstep admission pauses at a common scheduling boundary. WavePP overlaps preparation with earlier waves and materializes locally before E. Orange includes all CPU preparation, including materialization. Durations are schematic.}
    \label{fig:design-overlap}
\end{figure}

Figure~\ref{fig:design-overlap} illustrates this overlap.

\subsection{Fast Reuse Hints}
\label{sec:design-hints}

The naive implementation first walks each cache tree to find reusable state, then walks it again after agreement to pin the common prefix. Both walks require a mutex due to concurrent modifications. We could instead pin blocks during the first walk, but each rank would claim its local prefix before knowing how much the other ranks can reuse. Ranks with more cached state would then need to release blocks beyond the common endpoint. For recurrent caches, lowering the endpoint may also require acquiring an earlier snapshot and releasing the one just claimed. Across many waiting requests, these early claims reduce what the cache can evict to make room for requests ready to execute. Similarly, reserving space for suffixes early would tie up more memory, including for requests that may wait or be cancelled. Claiming only for the request at the head of the queue avoids much of this pressure, but leaves reuse discovery and agreement to be done when that request is ready for admission.

In order to reduce the overhead of repeated traversals and mutex contention, we build an index that provides a reuse hint to each rank. This index does not contain any blocks and is only updated on cache tree changes. Queued requests can therefore discover reuse and agree on a candidate without pinning cache or reserving suffix capacity. Only requests selected for admission then perform the authoritative walk, validating and pinning no more than the candidate prefix. This lets candidate agreement proceed before the request reaches the head of the queue, while bounding the later walk on each rank.

The common hint also limits how far each rank must search. Consider a surge of 200K-token requests where most stages retain nearly the whole prefix, but one stage can resume only at 32K for one of the requests. If 32K is usable on every stage, other cache rich stages need to walk and pin only 32K tokens of the prefix. Requests outside the admission limit can still obtain hints without taking ownership. If they are cancelled while waiting, there are no cache references or capacity reservations to undo.

\begin{figure}[!htbp]
    \centering
    \includegraphics[width=\linewidth]{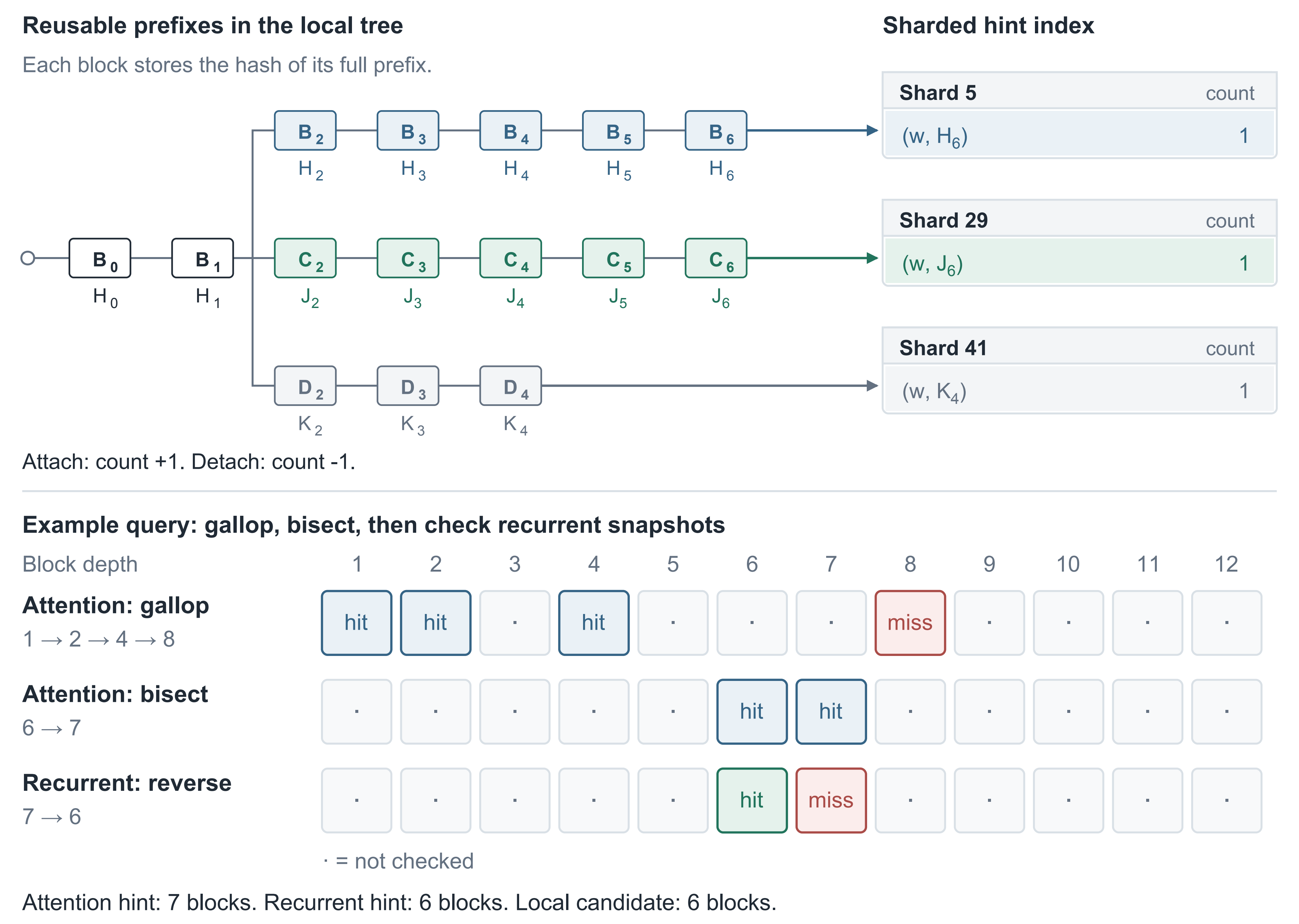}
    \caption{The sharded hint index and an example lookup. Attention probes gallop and bisect to seven blocks. The reverse recurrent scan lowers the local candidate to six. The later lease validates and pins the actual state.}
    \label{fig:design-hints}
\end{figure}

For each full token block $B_i$, we compute a chained prefix hash
\begin{equation}
    H_{-1}=0,\qquad H_i=\operatorname{hash}(B_i,\mathrm{extraKeys}_i,\mathrm{salt},H_{i-1}).
    \label{eq:design-hash}
\end{equation}
The parent hash thus makes the key dependent on the prefix and the current block. Each rank maintains an index keyed by the cache window and $H_i$, with a counter for each key. When a block is added to the tree, we increment its key's counter, and once it is detached from the tree, we reduce the counter by one, evicting the index entry if the count reaches zero. Offloading a block to a different storage method does not affect this index since offloading still preserves reuse.

Reading the hint index still requires acquiring the mutex of each queried shard. For dense attention prefixes, we reduce the number of probes from $O(N)$ to approximately $O(\log N)$ for $N$ eligible blocks by probing depths at increasing powers of two until the first miss, then bisecting between the last match and miss. Request hashing still requires $O(N)$ steps.

Recurrent snapshots are sparse, so we instead scan backward from the attention-limited endpoint. This scan can require $O(N)$ probes of the index, but typically requires far fewer. We take the minimum of the required local cache endpoints, then the all-rank minimum, to obtain the reuse candidate. This is still a hint, since lowering the endpoint may require finding an earlier recurrent snapshot.

Note that the index can be changed during lookup as the lock is not constantly held, so an authoritative final walk is still required (next section). Hence, the index provides a useful, but occasionally stale, view of the cache state to quickly enable ranks to agree upon a candidate reuse endpoint.

In a 200K-token, concurrency-64 stress experiment, the optimized sharded lookup reduced median probe time from 64.4~ms to 24~$\mu$s, and hint discovery reached a p99 of 2.35~ms.

\subsection{Local Leases}
\label{sec:design-leases}

\begin{figure}[!htbp]
    \centering
    \includegraphics[width=\linewidth]{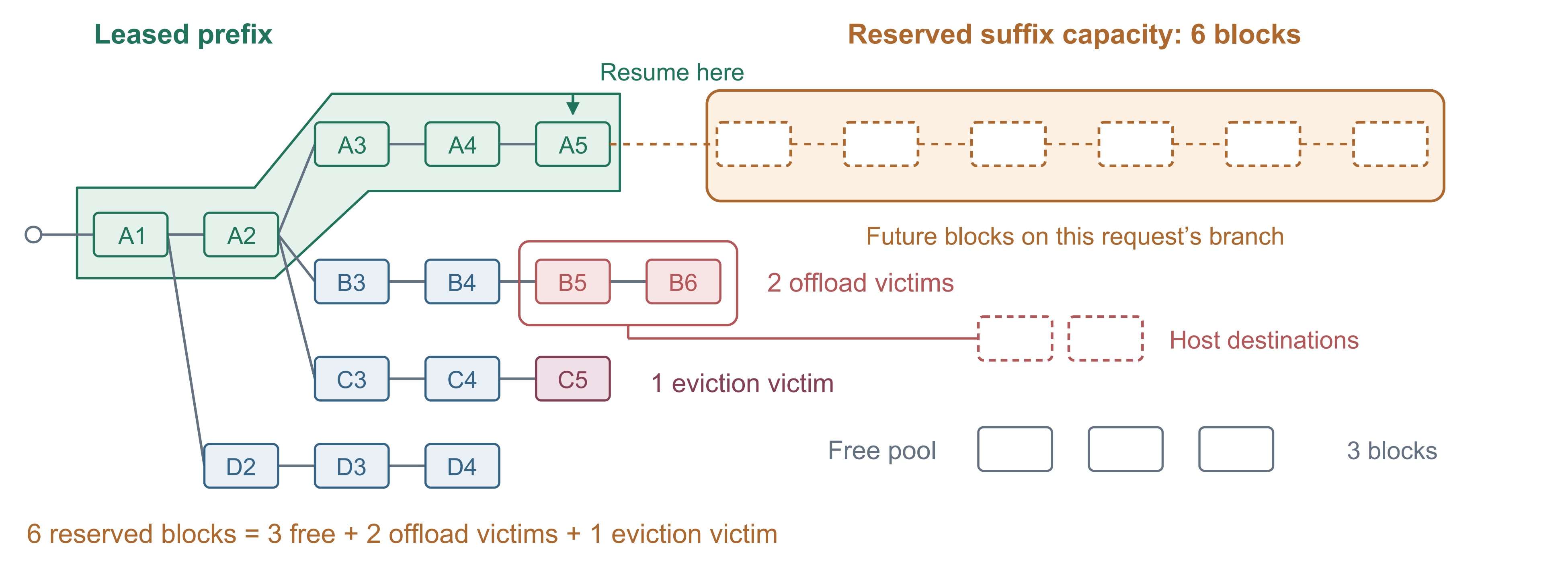}
    \caption{The lease protects the reused prefix and its final recurrent snapshot. Six blocks of suffix capacity are backed by three free blocks, two offload victims with host destinations, and one eviction victim. The dashed suffix has not yet been allocated.}
    \label{fig:design-ownership}
\end{figure}

After a candidate reuse endpoint is agreed upon, the cache must be pinned so that it cannot be evicted while the request awaits materialization. A \emph{local lease} maintains these references without physically installing the sequence (Figure~\ref{fig:design-ownership}). The admission thread builds the exact block keys outside the tree mutex, then walks the tree under mutex, and pins the required blocks based on reuse agreement. Other requests can also share the pinned state, but cannot evict it until the lease is released. The mutex is released before communicating the lease acquisition to reduce contention.

Once the lease is acquired, each rank reports the reuse endpoint it actually pinned. If endpoints differ, richer ranks trim their claims to a common endpoint before committing, or all ranks release the attempt and prepare again at the lower endpoint if any rank has already committed. In our regular experiments, we did not observe this discrepancy, but encountered it once under extreme load with frequent request cancellations when lease creation and abort churn was high.

For attention blocks, lowering the endpoint simply unpins the excess blocks, however, the next usable recurrent endpoint may once again diverge across the ranks. For example, suppose rank 0 has snapshots at 64K, 72K, and 90K, while rank 1 has them at 64K and 80K. At a candidate of 80K, rank 0 falls back to 72K. Rank 1 has no snapshot at 72K and must then fall back to 64K, which both ranks can use. Each adjustment acquires the earlier recurrent state before releasing the later one. Further reconciliation can only lower the endpoint. This process is bounded since every reconciliation monotonically decreases the claimed endpoint, ensuring eventual convergence.

Note that pinning reduces space for other requests to claim space for uncached suffixes. We therefore acquire leases in first-come, first-served (FCFS) ready order, and limit the number of requests that can hold leases simultaneously to prevent too many requests pinning cache and stalling the pipeline due to no eviction space. This is based on the number of admitted tokens required as well as remaining token capacity in the cache. We first cap the request count at the maximum batch size, since each wave may contain several requests. Secondly, we leave a token headroom proportional to $2PM$, where $P$ is the number of pipeline stages, and $M$ is the maximum number of tokens per wave. The exception to this rule is allowing a single oversized request to ignore the token headroom limit. This avoids blocking a single large request while still protecting the cache from being overwhelmed by many smaller requests.

The lease has no expiry based on time and is held through materialization until the sequence acquires its own references. This leaves no gap in ownership.

\subsection{Capacity Escrow}
\label{sec:design-escrow}

A leased prefix guarantees reuse, but the uncached suffix also needs space. If we admit requests on prefix hits alone, they can reserve most of the cache but leave too little room for any of them to finish. We therefore reserve cache space for each layer (a \emph{capacity escrow}) immediately after each local lease. If this fails, then the lease is released and the request waits until sufficient capacity is available.

The reservation is first prepared using the rank's local lease endpoint. An \emph{escrow commit} finalizes the reservation at a fixed reuse endpoint. Commit rechecks the required suffix capacity and the reserved victims, acquires any additional or replacement capacity, and marks the reservation as committed. It does not yet allocate the request's blocks or evict the victims. Those operations remain part of local materialization.

Available capacity includes both free blocks and blocks that can be reclaimed by offloading or evicting cache. We follow the allocator's existing eviction order. When we require offloading a victim, we reserve space for the destination before acquiring the primary block. The escrow holds concrete references to victim blocks, but leaves the contents attached to the tree without evicting or copying them. The actual eviction or offloading steps are deferred to the local executor during materialization, allowing the reservation work to overlap GPU execution. Recurrent placeholders are interchangeable and hence reserved by count rather than by holding individual snapshot destinations. Materialization uses the production allocator while the escrow remains held, rather than allocating directly from its reserved block list.

Since we leave victim blocks in the tree, other requests can still claim them for reuse. At commit, the escrow detects the additional owner through the block's reference count and selects a replacement from the same cache window. It acquires the replacement, including any required host destination, before releasing its reference to the original victim. The claiming request keeps the original content alive, while the replacement backs the reserved suffix capacity. The admission limits and capacity floors retain headroom for active requests and restrict how much cache pending leases can pin. This headroom is intended to accommodate victim substitution while earlier requests finish. Commit rechecks the capacity floors before publishing the substitutions.

Committing a capacity escrow also recomputes demand if the reuse endpoint has reduced. For example, reducing reuse from 112K to 96K in a 200K-token prompt increases the uncached suffix from 88K to 104K. The rank must reserve additional capacity before committing the lower endpoint. The same capacity checks apply when replacing a victim. Commit gathers additional capacity and victim replacements under the tree mutex, then publishes the revised reservation only if all checks succeed. A failure releases only the new acquisitions, leaving the original reservation intact for distributed cleanup.

Commit can run immediately after preparation when the exact lease preserves the hint candidate. Each rank reports the result with its lease vote, avoiding a separate commit round if all ranks succeed at that endpoint. Once all ranks report successful commits at the same endpoint, rank 0 can schedule after its own materialization without another round of enqueue acknowledgements. If the preparation, commit, or enqueue fails, every rank releases the escrow first, then the lease. Otherwise, each rank holds its escrow and lease until its sequence owns the real cache blocks, then releases the escrow followed by the lease.

\subsection{Pipeline Scheduling}
\label{sec:design-scheduling}

\begin{figure}[!htbp]
    \centering
    \includegraphics[width=\linewidth]{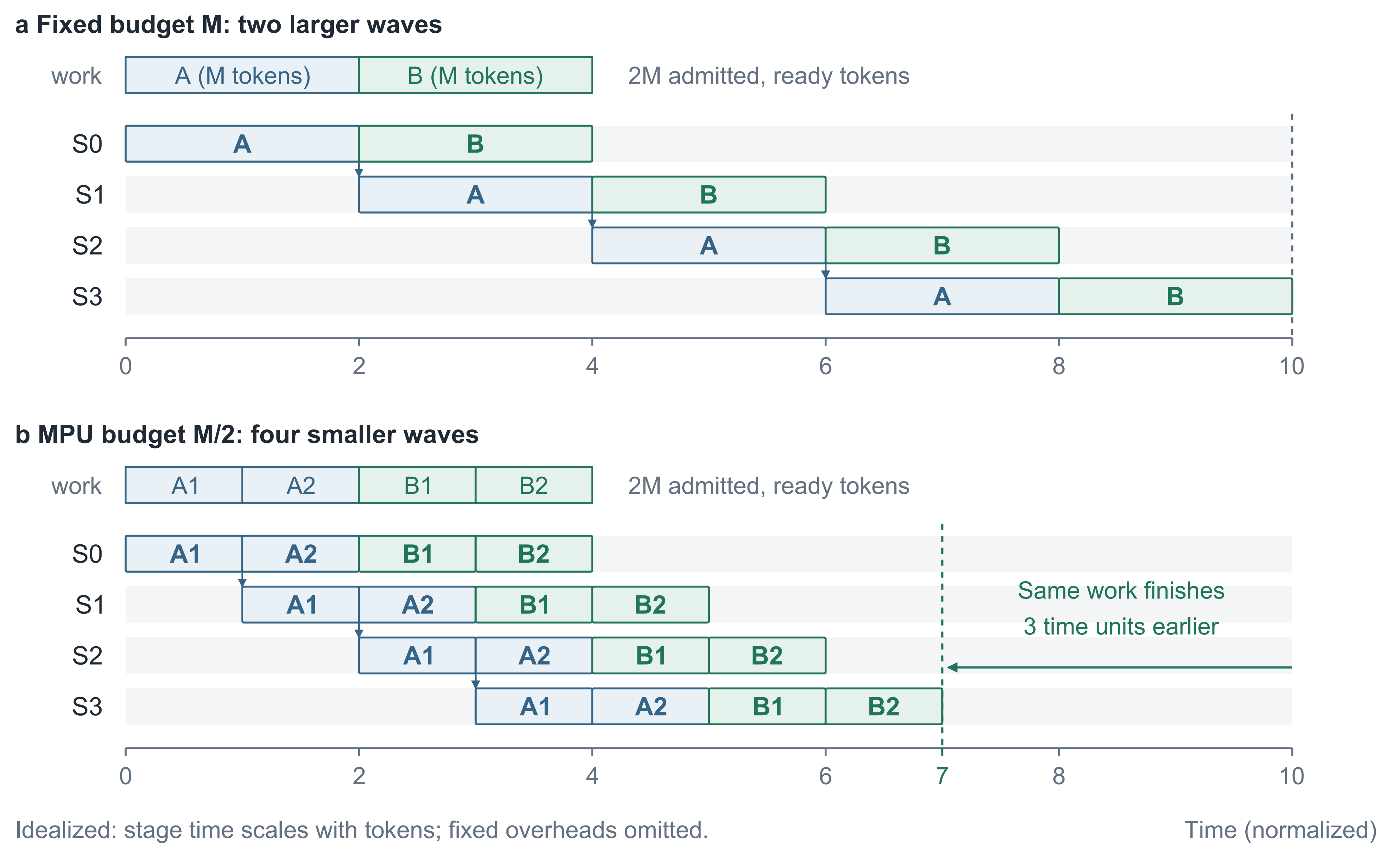}
    \caption{Wave sizing with the same $2M$ tokens on four stages. A budget of $M$ gives two waves, while $M/2$ gives four. Stage time is assumed proportional to token count and fixed overheads are omitted, so this illustrates the scheduling effect rather than a measured speedup.}
    \label{fig:design-mpu}
\end{figure}

Even when admission maintains a fast cadence, a purely greedy scheduling approach often leads to suboptimal pipeline utilization. Consider a set of requests with a combined $2M$ tokens, where $M$ is the pipeline per-wave budget. If packed into full waves, this creates only two waves in total, and at any given time, at most two ranks process requests, leaving other ranks idle and the pipeline underutilized. Although many existing approaches for dynamically chunking requests exist, we adopt a very simple strategy by noticing that halving the budget in this case leads to 4 waves, as Figure~\ref{fig:design-mpu} shows. Smaller waves also add overhead due to additional forward launches and activation transfers, so we only shrink the budget of waves when little admitted work remains. We use Maximum Pipeline Utilization (MPU), a variant of prefill throttling~\citep{guo_gllm:_2025} tailored to our PP design. MPU decides the wave budget to keep the pipeline full, accounting for admitted work and chunks already in flight. Cache capacity is handled separately during admission.

The executor maintains an execution ring of $R=2P$ waves, allowing one wave to execute and another to wait at each stage. Rank 0 can therefore schedule further work before the result of every earlier wave returns. To keep this ring supplied, the planner targets $S=R+(F-1)P$ waves, where $F-1$ adds waiting pipeline fills. The matched PP4 experiments use $F=2$.

The planner chooses from a sequence that repeatedly halves the maximum wave budget $M$ down to a specified floor $b_{\min}$:
\begin{equation}
    b_0=M,\qquad
    b_{j+1}=\max\!\left(b_{\min},\operatorname{alignDown}_{B}(b_j/2)\right).
    \label{eq:design-budget}
\end{equation}
Here $B$ is the block size, and $M$ and $b_{\min}$ are block-aligned with $B\leq b_{\min}\leq M$. We stop halving at $b_{\min}$. Let $W$ include the not-yet-scheduled suffix tokens of admitted requests and their most recently committed chunks still occupying the ring. Counting this in-flight work avoids shrinking the budget simply because earlier chunks have left the planner.

Token count alone can underestimate how many waves are available. Two request pieces larger than half a wave cannot share that wave. For a candidate budget $b$, let $N_{>b/2}$ count such pieces among the remaining suffixes and in-flight chunks. We estimate the available waves as
\begin{equation}
    n(b)=\max\!\left(W/b,\,N_{>b/2}\right).
\end{equation}
We select the largest budget with $n(b)\geq S$, falling back to $b_{\min}$ if none qualifies. Once a larger budget already supplies $R$ waves, further thinning is limited when it would mainly split requests that fit the larger budget. In particular, pieces that would be newly split must account for at most 12.5\% of $W$. This avoids adding forward launches and transfers just to create more chunks of the same requests.

Large bursts can increase the budget immediately. A one-step increase must persist for $R/2$ planning decisions, while a decrease must persist for $R$ decisions and pass a hysteresis check. The number of lanes depends on the maximum batch size, so changing the wave budget does not remove an existing request's lane. During packing, we also avoid filling a small leftover space with a fragment of a larger request when that fragment is smaller than $\max(b_{\min},b/2)$. The matched PP4 comparison uses $M=16{,}384$ and $b_{\min}=8{,}192$. The Kimi K3 scheduling settings are given in Section~\ref{sec:eval-setup}.

Rank 0 is responsible for packing the requests in FCFS order. While the ring is busy, a lone wave below half capacity may wait one iteration for another request in admission so they can share a wave. The scheduler still checks resources and may accept less work, so the planner advances its token positions only by the spans actually accepted. Followers cannot infer the selected chunks from their local cache views. Therefore, rank 0 sends each scheduled chunk's request ID, starting token, and length to other ranks. Followers apply those spans after local materialization while preserving both request order and token boundaries. If a position mismatch remains after materialization, execution fails rather than processing activations against the wrong cache state. This keeps scheduling consistent, while cache management proceeds independently across all stages.

Successive context chunks can enter while earlier chunks remain in downstream stages. Their order is preserved on each stage's execution stream.

\renewcommand{\floatpagefraction}{0.85}
\section{Evaluation}
\label{sec:evaluation}

\makeatletter
\setlength{\@fptop}{0pt}
\setlength{\@fpsep}{12pt plus 2pt minus 2pt}
\makeatother

\definecolor{EvalBestFill}{HTML}{E5F3E9}
\captionsetup[table]{font=footnotesize,width=\textwidth,margin=0pt,skip=7pt}
\newcommand{\evaltablesetup}{%
  \fontsize{9}{11}\selectfont
  \renewcommand{\arraystretch}{1.15}%
  \setlength{\tabcolsep}{4pt}%
  \arrayrulecolor{black}%
  \setlength{\arrayrulewidth}{0.4pt}%
}
\newcommand{\evalbest}[1]{%
  \cellcolor{EvalBestFill}%
  \underline{#1}%
}
\newcommand{\evalsecond}[1]{\underline{#1}}

We evaluate WavePP on cold prefill, requests with a shared cached prefix, and mixtures of the two. We first measure how WavePP improves prefill throughput within TensorRT-LLM PP4 on GLM~5.2 and MiniMax M2.7, using the same kernels as concurrency and cache pressure increase. We then compare PP8 and TP8 across TRT-LLM, SGLang, and vLLM on 28 Kimi K3 workload and concurrency settings. Finally, we examine prefix retention and admission and scheduling ablations.

\subsection{Metrics}
\label{sec:eval-metrics}

We report \textbf{aggregate prompt throughput}, measured as the total input tokens of completed requests divided by the run's wall time. This includes cached tokens, so it measures the input served rather than only the tokens computed. We use $c$ for request concurrency. We also report \textbf{prefill completion time}, the time taken to complete the entire prefill of a request, including admission and queueing. For the Kimi K3 comparisons, we report p50 and p95 across requests.

\paragraph{Table conventions}
Green shading and an underline mark the best overall value. Underlining alone marks the second-best PP value and the best TP value, or the second-best TP value when TP has the best overall value. Ties at the reported precision share the same marking.

In the Kimi K3 tables, $\Delta_{\mathrm{TP}}$ and $\Delta_{\mathrm{PP}}$ compare WavePP with the best TP and external PP result, respectively, for each metric. Both use $(\mathrm{WavePP}/\mathrm{baseline}-1)\times100\%$. Positive throughput deltas indicate higher throughput for WavePP, and negative latency deltas indicate lower completion time. Latency deltas use the rounded times shown in the tables. Retention differences are in percentage points.

\subsection{Pipeline Parallelism in TensorRT-LLM}
\label{sec:eval-trt-pp}

We first examine how WavePP's admission and scheduling changes compare to TensorRT-LLM's original PP runtime. We compare TensorRT-LLM PP4 and WavePP4 on GLM~5.2 \citep{glm52_model_card,glm-5-team_glm-5:_2026} and MiniMax M2.7 \citep{minimax27_model_card,chen_minimax-m2_2026}, using the same TensorRT-LLM 1.3.0rc26 kernels and model configurations. This measures the combined runtime changes within a four-stage pipeline.

\paragraph{Setup}
Both models use NVFP4 weights and four B300 GPUs per system. We tune the TensorRT-LLM PP4 configuration to a 16,384-token microbatch limit and use the same maximum wave budget for WavePP. Both systems allow up to 128 requests per batch. WavePP uses its standard configuration, including the $2P$ execution ring, $F=2$, and an 8,192-token minimum wave budget. Each system keeps one configuration across all workloads, with diagnostic instrumentation disabled. We report medians from three independent server boots per system and model, with the runs interleaved on the same node. The rest of the configuration is identical across systems including the pipeline partition.

Cold workloads run first within each boot. Subsequent reuse workloads therefore encounter a cache pool that has already filled and evicted earlier prompts. This tests admission during continued use of the cache. We include short and long cold inputs, 90\% prefix reuse at both lengths, and mixed traffic with 75\% warm short requests and 25\% cold long requests. The recorded short/long lengths are approximately 128K/260K for GLM and 129K/195K for MiniMax. A further workload uses a 4,096-token uncached suffix behind a shared prefix of about 125K tokens and extends concurrency to 128. Throughput counts the full input, including the cached prefix. Table~\ref{tab:trt-pp-complete} reports all 40 settings. The goal of the short suffix workload is to stress the runtime's admission mechanisms under high cache pressure and rapid request turnover.

\begin{figure}[!htbp]
  \centering
  \includegraphics[width=\linewidth]{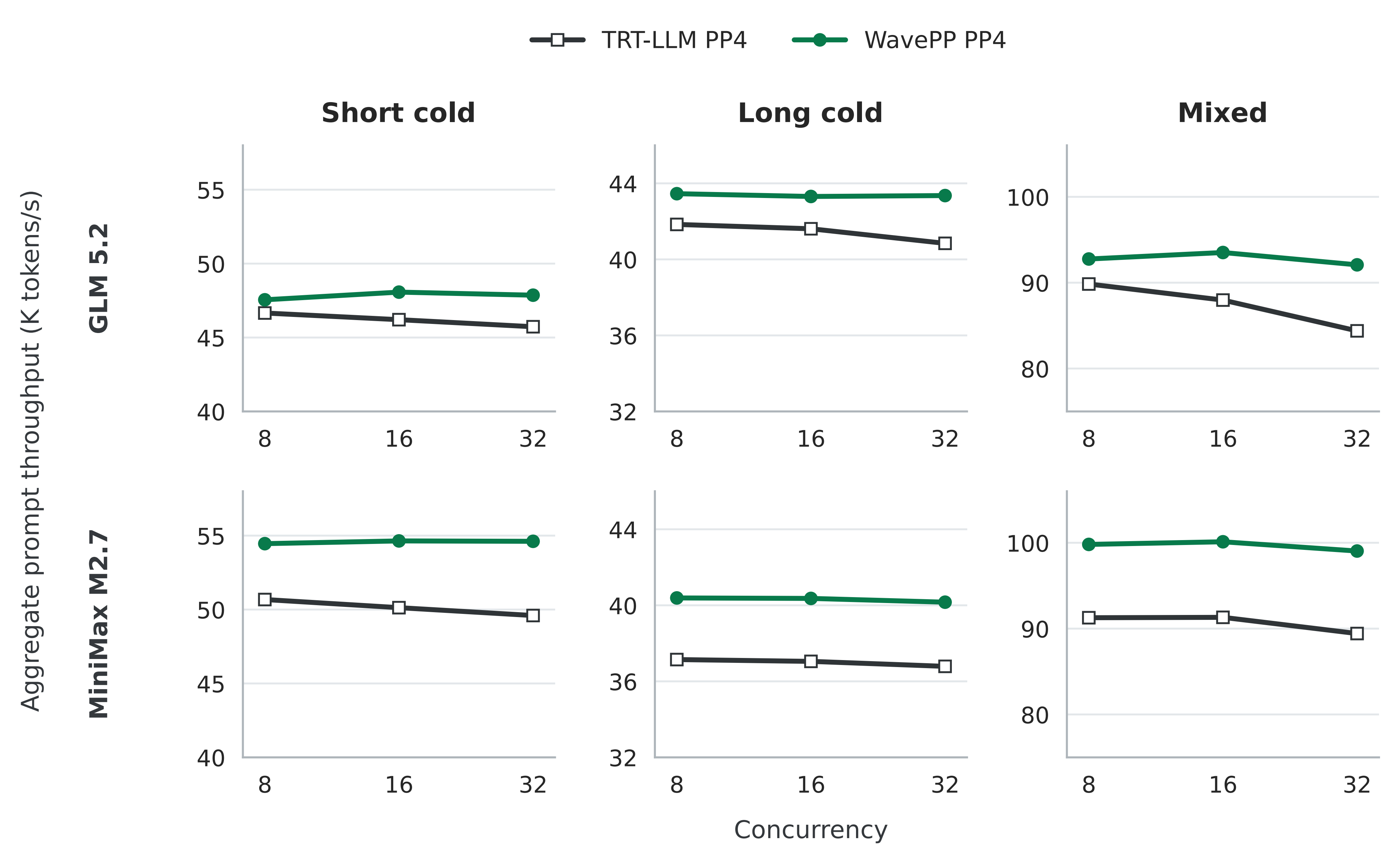}
  \caption{Prefill throughput with TensorRT-LLM PP4 and WavePP4 on the same kernels. Cold and mixed workloads use short/long inputs of approximately 128K/260K tokens for GLM and 129K/195K for MiniMax. Mixed traffic is 75\% warm short requests and 25\% cold long requests. Points are median aggregate prompt throughput across three server boots.}
  \label{fig:trt-pp-cold}
\end{figure}

\paragraph{Throughput under sustained load}
The admission and scheduling changes improve throughput relative to TensorRT-LLM PP4 in 37 of 40 settings, including every cold and mixed setting. On cold and mixed traffic, WavePP improves throughput by 1.9--9.1\% for GLM and 7.5--10.7\% for MiniMax (Figure~\ref{fig:trt-pp-cold}). With 90\% reuse of the shorter inputs, WavePP improves throughput at $c=16$ and $c=32$ by 11.5\% and 17.8\% for GLM, and 9.1\% and 10.4\% for MiniMax.

\begin{figure}[!htbp]
  \centering
  \includegraphics[width=\linewidth]{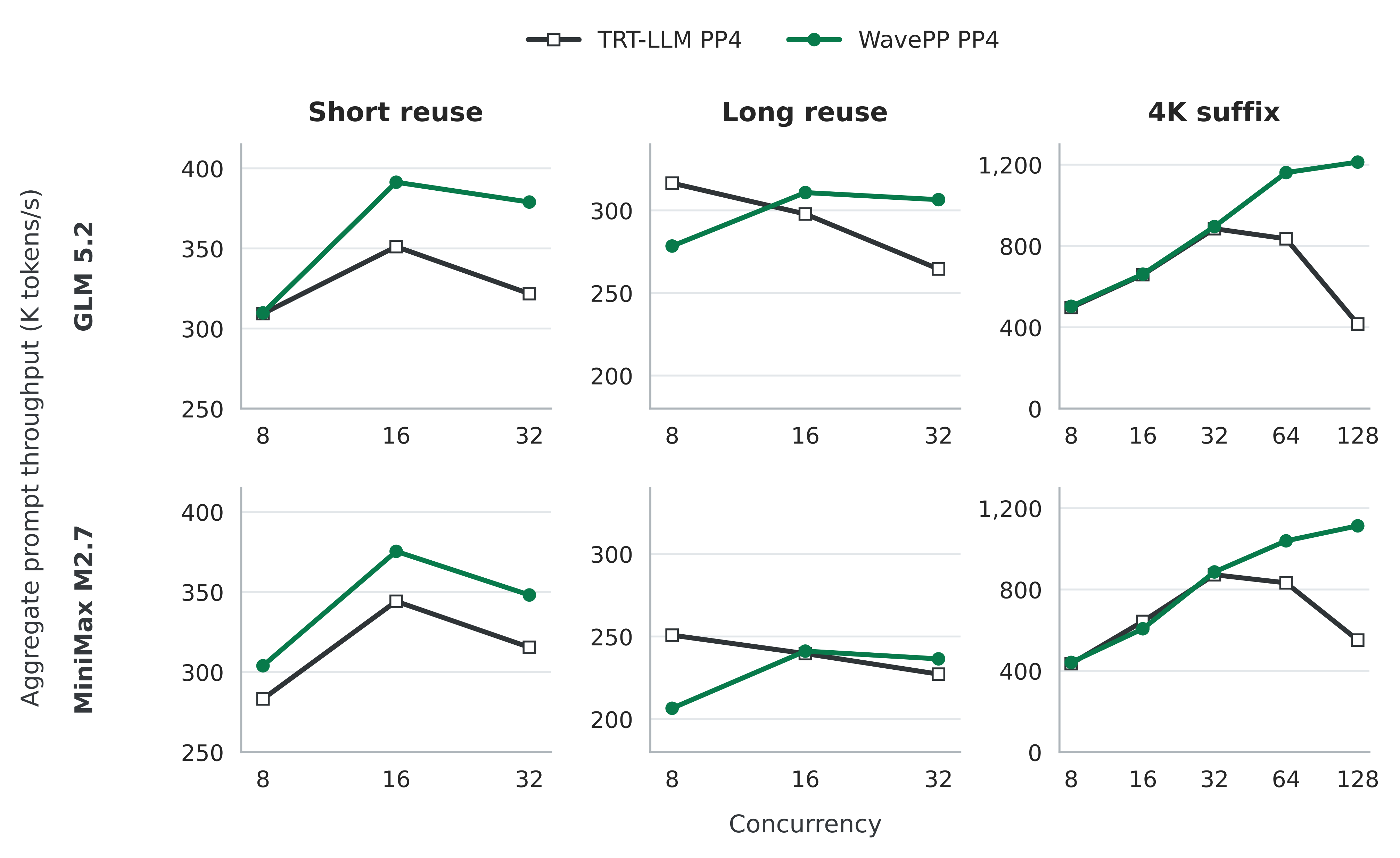}
  \caption{Reuse throughput with the same kernels and PP4 topology. The first two columns reuse 90\% of the short and long inputs. The last column uses a 4,096-token suffix and extends to $c=128$. Reuse follows cold workloads within each server boot, so these measurements include cache pressure from earlier traffic. Throughput includes cached input tokens.}
  \label{fig:trt-pp-reuse}
\end{figure}

The largest throughput improvements appear when many requests reuse a long prefix and compute only a short suffix (Figure~\ref{fig:trt-pp-reuse}). These requests complete quickly, increasing the rate at which the runtime must prepare and schedule new work. At $c=64$, WavePP improves throughput by 39.0\% on GLM and 24.8\% on MiniMax. At $c=128$, it sustains \num{1212572} and \num{1113196} input tok/s, compared with \num{416294} and \num{551143} for TensorRT-LLM PP4. WavePP provides $2.91\times$ and $2.02\times$ the throughput of TensorRT-LLM PP4 after the cache has been filled by earlier cold traffic.

For the 4K-suffix workload at $c=128$, median prefill completion time also falls by approximately 71\% on GLM and 55\% on MiniMax. However, throughput is lower in three settings: long-prefix reuse at $c=8$ for both models and the 4K-suffix MiniMax workload at $c=16$, as shown in the figure and table.

\begin{table}[!htbp]
  \centering
  \caption{PP4 comparison on matched kernels. Aggregate prompt throughput
    (tokens/s, including cached tokens) is the median of three boots per
    system and model. Each system uses one configuration throughout.
    Throughput change $\Delta$ is $(\mathrm{WavePP}/\text{TRT-LLM}-1)\times100\%$.
    Green shading marks the higher throughput. Short and long reuse use
    90\% cached prefixes. The 4K-suffix
    cells follow cold traffic and retain its cache-pool eviction history.}
  \label{tab:trt-pp-complete}
  \evaltablesetup
  \begin{tabular*}{\textwidth}{@{\extracolsep{\fill}}lrrrrrrr@{}}
    \toprule
    & & \multicolumn{3}{c}{GLM 5.2} & \multicolumn{3}{c}{MiniMax M2.7} \\
    \cmidrule(lr){3-5}\cmidrule(lr){6-8}
    Family & $c$ & TRT-LLM & WavePP & $\Delta$ & TRT-LLM & WavePP & $\Delta$ \\
    \midrule
    Short cold & 8 & \evalsecond{46,658} & \evalbest{47,552} & +1.9\% & \evalsecond{50,677} & \evalbest{54,457} & +7.5\% \\
     & 16 & \evalsecond{46,207} & \evalbest{48,071} & +4.0\% & \evalsecond{50,127} & \evalbest{54,642} & +9.0\% \\
     & 32 & \evalsecond{45,730} & \evalbest{47,864} & +4.7\% & \evalsecond{49,596} & \evalbest{54,615} & +10.1\% \\
    \midrule
    Long cold & 8 & \evalsecond{41,839} & \evalbest{43,456} & +3.9\% & \evalsecond{37,146} & \evalbest{40,388} & +8.7\% \\
     & 16 & \evalsecond{41,610} & \evalbest{43,310} & +4.1\% & \evalsecond{37,051} & \evalbest{40,363} & +8.9\% \\
     & 32 & \evalsecond{40,844} & \evalbest{43,357} & +6.2\% & \evalsecond{36,789} & \evalbest{40,165} & +9.2\% \\
    \midrule
    Short reuse & 8 & \evalsecond{309,301} & \evalbest{309,762} & +0.1\% & \evalsecond{283,199} & \evalbest{303,921} & +7.3\% \\
     & 16 & \evalsecond{351,119} & \evalbest{391,338} & +11.5\% & \evalsecond{344,174} & \evalbest{375,368} & +9.1\% \\
     & 32 & \evalsecond{321,750} & \evalbest{378,951} & +17.8\% & \evalsecond{315,446} & \evalbest{348,107} & +10.4\% \\
    \midrule
    Long reuse & 8 & \evalbest{316,516} & \evalsecond{278,418} & -12.0\% & \evalbest{250,815} & \evalsecond{206,548} & -17.6\% \\
     & 16 & \evalsecond{297,822} & \evalbest{310,786} & +4.4\% & \evalsecond{239,657} & \evalbest{241,150} & +0.6\% \\
     & 32 & \evalsecond{264,520} & \evalbest{306,497} & +15.9\% & \evalsecond{227,220} & \evalbest{236,459} & +4.1\% \\
    \midrule
    Mixed & 8 & \evalsecond{89,844} & \evalbest{92,759} & +3.2\% & \evalsecond{91,270} & \evalbest{99,810} & +9.4\% \\
     & 16 & \evalsecond{87,974} & \evalbest{93,515} & +6.3\% & \evalsecond{91,314} & \evalbest{100,117} & +9.6\% \\
     & 32 & \evalsecond{84,387} & \evalbest{92,080} & +9.1\% & \evalsecond{89,427} & \evalbest{99,039} & +10.7\% \\
    \midrule
    4K suffix & 8 & \evalsecond{497,230} & \evalbest{503,184} & +1.2\% & \evalsecond{435,088} & \evalbest{441,314} & +1.4\% \\
     & 16 & \evalsecond{658,662} & \evalbest{661,230} & +0.4\% & \evalbest{642,896} & \evalsecond{606,764} & -5.6\% \\
     & 32 & \evalsecond{884,810} & \evalbest{895,482} & +1.2\% & \evalsecond{872,496} & \evalbest{886,305} & +1.6\% \\
     & 64 & \evalsecond{835,235} & \evalbest{1,160,931} & +39.0\% & \evalsecond{832,645} & \evalbest{1,039,376} & +24.8\% \\
     & 128 & \evalsecond{416,294} & \evalbest{1,212,572} & +191.3\% & \evalsecond{551,143} & \evalbest{1,113,196} & +102.0\% \\
    \bottomrule
  \end{tabular*}
\end{table}

The smaller throughput improvements on cold inputs in Figure~\ref{fig:trt-pp-cold} are consistent with the greater amount of GPU computation per request. Each admission supplies many chunks, giving both runtimes more work to execute before another request must be prepared. Reducing admission and scheduling delays therefore affects a smaller fraction of the total processing time.

The 4K-suffix results show a different trend. From $c=32$ to $c=128$, TensorRT-LLM PP4 throughput decreases for both models, while WavePP throughput increases (Table~\ref{tab:trt-pp-complete}). More concurrent requests compete for space in the filled cache pool, while each admission supplies little computation. Preparing reuse and reserving suffix capacity during earlier execution helps WavePP keep new requests ready as others finish. The ablation in Section~\ref{sec:eval-ablation} separately examines admission and scheduling on Kimi K3.

\subsection{Kimi K3 Experimental Setup}
\label{sec:eval-setup}

\paragraph{Model and hardware}
We next evaluate WavePP across serving libraries on Kimi K3, a 2.8-trillion-parameter hybrid KDA/MLA MoE model with both recurrent state and attention caches \citep{team_kimi_2026}. These cross-library experiments use the same eight GB300 GPUs arranged as two four-GPU compute nodes. WavePP and the pipeline baselines use TP1$\times$PP8 and the tensor baselines use TP8/EP8.

\paragraph{Systems}
WavePP is implemented in TensorRT-LLM \citep{tensorrt_llm}. We compare it with TRT-LLM TP8/EP8, and with the TP8/EP8 and PP8 implementations in SGLang and vLLM (\Cref{tab:eval-baselines}). WavePP and the TRT-LLM TP8/EP8 reference share the same Kimi K3 kernels and library. Both SGLang configurations use v0.5.18 at commit \texttt{71de97b264b0} and the same hybrid host-cache configuration \citep{sglang_repository}.

WavePP reserves 128~GiB of host cache per rank. Its scheduler uses a 16,384-token limit for both total work and context chunks. SGLang uses a 16K prefill budget, and vLLM PP8 uses a 16K packed-token budget (\Cref{tab:eval-baselines}). For Kimi K3, MPU uses the remaining unscheduled tokens to choose between two budgets, $M$ and $M/2$, targeting one pipeline fill.

As a separate baseline check, our SGLang PP8 run with 8K cold inputs reaches approximately 6,300 input tokens/s per GPU on two four-GPU GB300 nodes. The SGLang day-0 report highlights 5,958 input tokens/s per GPU for 8K prefill at $c=64$ on the same topology \citep{sglang_kimi_k3_day0}. Our result is approximately 5.7\% higher.

\begin{table}[!htbp]
  \centering
  \caption{Configurations of the Kimi K3 comparison baselines.}
  \label{tab:eval-baselines}
  \evaltablesetup
  \begin{tabularx}{\linewidth}{@{}>{\raggedright\arraybackslash}p{0.16\linewidth}
    >{\raggedright\arraybackslash}p{0.22\linewidth}X@{}}
    \toprule
    Baseline & Parallel configuration & Benchmark configuration \\
    \midrule
    vLLM TP8/EP8 & TP8/EP8 on eight GPUs & v0.28.0 with the FlashInfer
      TensorRT-LLM MoE path \\
    SGLang TP8/EP8 & TP8/EP8 on eight GPUs & Same SGLang build, 16K prefill
      budget, and hybrid host-cache configuration as the SGLang PP8 arm \\
    TRT-LLM TP8/EP8 & TP8/EP8 on eight GPUs & Same Kimi K3 engine family and shared
      kernel changes as WavePP; valid two-node tensor-parallel placement \\
    \midrule
    vLLM PP8 & Eight-stage pipeline parallelism & v0.28.0 with a 16K
      packed-token budget and prefix caching enabled \\
    SGLang PP8 & Eight-stage pipeline parallelism & 16K prefill chunks;
      128~GiB/rank hybrid MLA+Mamba host cache; write-through offload, kernel
      I/O, and page-first loading \\
    \bottomrule
  \end{tabularx}
\end{table}

\paragraph{Workloads}
\Cref{tab:workload-grid} lists the workloads and request concurrency $c$. We use 65.5K, 131K, and 262K for 65,536, 131,072, and 262,144 input tokens, respectively. Seeded-reuse experiments first send standalone prefix requests to populate the cache. Mixed traffic contains 75\% warm 131K requests and 25\% cold 262K requests by request count. For 262K inputs, vLLM PP8 uses 262,080 tokens to leave room for the output token within its context limit, while the other systems use 262,144. Throughput uses the actual input token count.

All systems receive the same request sequence, subject to the context-length adjustment above. Request bodies are prepared before timing, and all requests in these Kimi K3 experiments completed successfully.

\begin{table}[!htbp]
  \centering
  \caption{Kimi K3 workloads and offered concurrency. Input lengths are abbreviated in decimal thousands; the vLLM PP8 262K exception is described in the text.}
  \label{tab:workload-grid}
  \evaltablesetup
  \begin{tabularx}{\linewidth}{@{}lrrXl@{}}
    \toprule
    Family & Input tokens & Shared prefix & Request composition & Concurrency \\
    \midrule
    Cold 65.5K & 65,536 & 0\% & cold & 1, 4, 8, 16, 32 \\
    Cold 131K & 131,072 & 0\% & cold & 1, 4, 8, 16, 32 \\
    Cold 262K & 262,144 & 0\% & cold & 1, 4, 8, 16, 32 \\
    Reuse 131K & 131,072 & 90\% & seeded shared prefix & 1, 4, 8, 16, 32 \\
    Reuse 262K & 262,144 & 90\% & seeded shared prefix & 1, 4, 8, 16, 32 \\
    Mixed & -- & -- & 75/25 warm-131K/cold-262K & 8, 16, 32 \\
    \bottomrule
  \end{tabularx}
\end{table}

\subsection{Kimi K3 Overall Performance}
\label{sec:eval-overall}

\begin{figure}[!htbp]
  \centering
  \includegraphics[width=\linewidth]{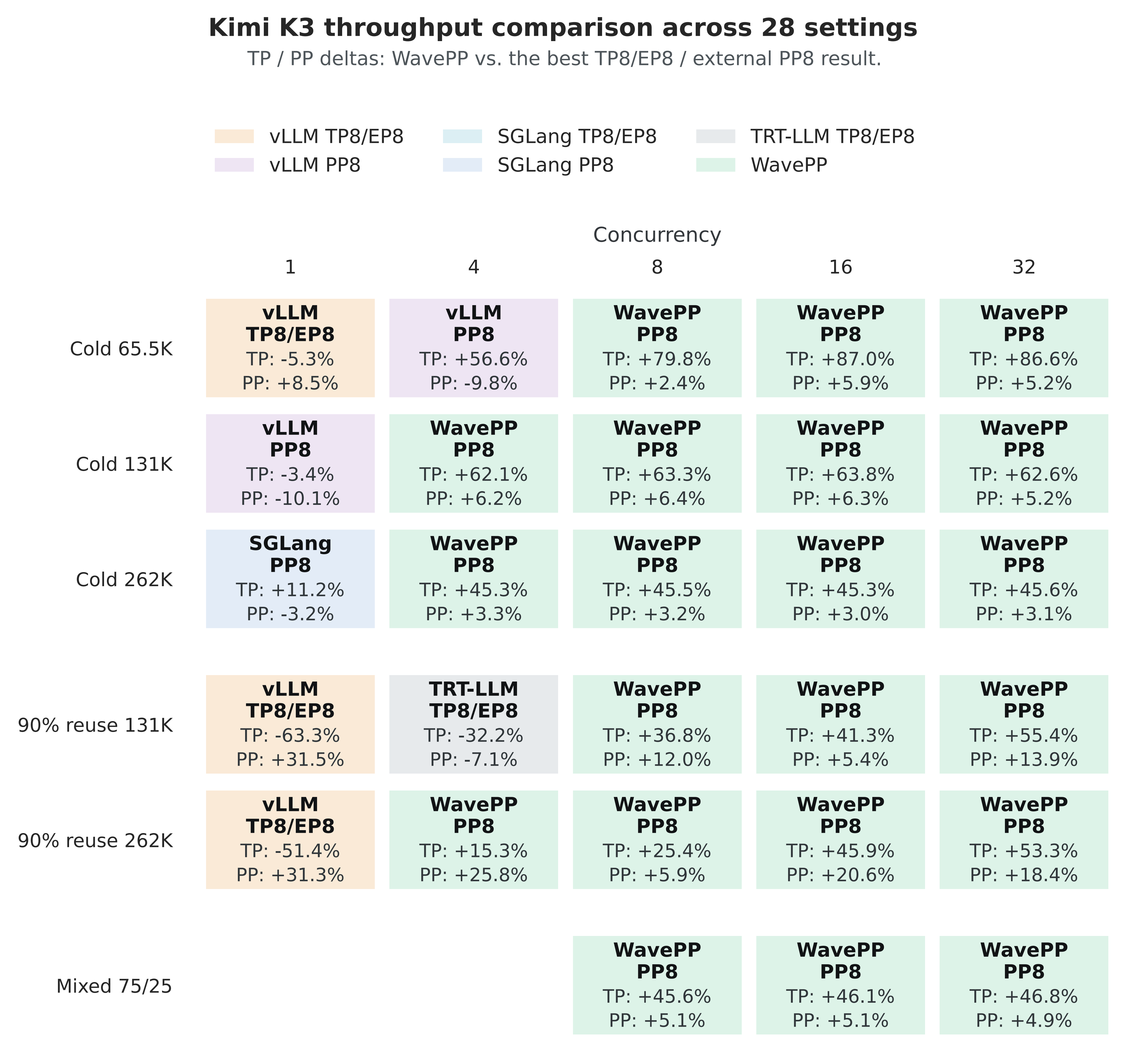}
  \caption{The system with the highest measured aggregate throughput in each Kimi K3 cell.  Each tile names the library with the highest throughput and reports WavePP's throughput difference relative to the highest-throughput TP8/EP8 and external PP8 results in that cell.  Positive values indicate higher throughput for WavePP. Color identifies only the library with the highest measured throughput.}
  \label{fig:overall-speedups}
\end{figure}

WavePP has the highest measured throughput in 21 of the 28 settings (Figure~\ref{fig:overall-speedups}), including all 18 settings at $c\geq8$. WavePP has lower throughput than some baselines in a few low concurrency settings when there is less independent work to fill the pipeline, though these gaps are also present for other PP methods as well.

For Kimi K3, the PP topology appears more favorable for prefill than TP. For example, SGLang PP8 improves geometric-mean throughput by 26.8\% over SGLang TP8/EP8. WavePP further improves geometric-mean throughput by 9.2\% over SGLang PP8 and 15.7\% over vLLM PP8 across the full grid.

\subsection{Cold Long-Context Prefill}
\label{sec:eval-cold}

Cold requests compute the full input without reusing earlier cache state. WavePP improves geometric-mean throughput over TRT-LLM TP8/EP8 by 58.8\%, 49.1\%, and 37.8\% for 65.5K, 131K, and 262K inputs, respectively. At $c=32$, it sustains \num{44643}, \num{35878}, and \num{28031} input tok/s.

The differences between PP implementations are smaller on these cold inputs, which require more GPU computation per request. Across the same three input lengths, WavePP improves geometric-mean throughput by 2.2\%, 4.2\%, and 8.0\% over vLLM PP8, and by 9.4\%, 4.5\%, and 1.9\% over SGLang PP8. WavePP benefits most at higher concurrencies. At $c=1$, WavePP's throughput is below the best baseline result at every input length, and its throughput remains below vLLM PP8 for cold 65.5K at $c=4$.

\begin{figure}[!htbp]
  \centering
  \includegraphics[width=\linewidth]{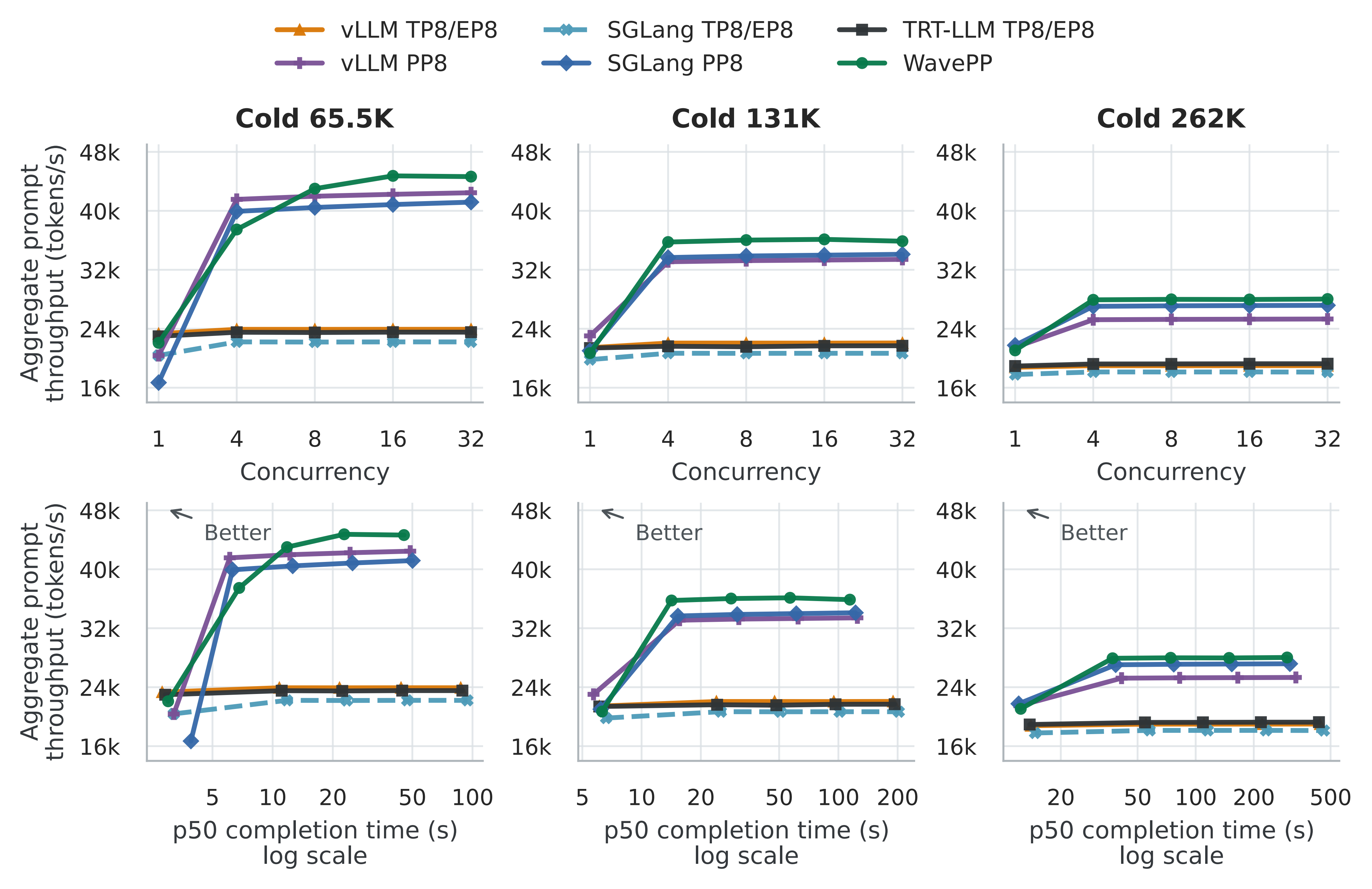}
  \caption{Cold long-context performance. The top row shows throughput by concurrency. The bottom row shows throughput p50 completion-time operating points. The arrow in each operating-point panel points toward higher throughput and lower completion time.}
  \label{fig:cold-performance}
\end{figure}

\begin{table}[!htbp]
  \centering
  \caption{Cold aggregate prompt throughput (input tok/s). Positive deltas indicate higher throughput for WavePP.}
  \label{tab:cold-throughput}
  \evaltablesetup
  \setlength{\tabcolsep}{2pt}
  \begin{tabular*}{\linewidth}{@{\extracolsep{\fill}}lrrr@{\hspace{3pt}}|@{\hspace{3pt}}rrrrr@{}}
    \toprule
    Context, $c$ & \shortstack{vLLM\\TP8/EP8} & \shortstack{SGLang\\TP8/EP8} &
      \shortstack{TRT-LLM\\TP8/EP8} & \shortstack{vLLM\\PP8} &
      \shortstack{SGLang\\PP8} & WavePP &
      $\Delta_{\mathrm{TP}}$ &
      $\Delta_{\mathrm{PP}}$ \\
    \midrule
    65.5K, 1 & \evalbest{23,330} & 20,378 & \evalsecond{22,976} & \evalsecond{20,361} & 16,691 & 22,089 & -5.3\% & +8.5\% \\
    65.5K, 4 & \evalsecond{23,922} & 22,215 & 23,524 & \evalbest{41,554} & \evalsecond{39,926} & 37,467 & +56.6\% & -9.8\% \\
    65.5K, 8 & \evalsecond{23,917} & 22,194 & 23,490 & \evalsecond{41,980} & 40,458 & \evalbest{43,007} & +79.8\% & +2.4\% \\
    65.5K, 16 & \evalsecond{23,925} & 22,215 & 23,537 & \evalsecond{42,242} & 40,852 & \evalbest{44,746} & +87.0\% & +5.9\% \\
    65.5K, 32 & \evalsecond{23,925} & 22,213 & 23,536 & \evalsecond{42,456} & 41,180 & \evalbest{44,643} & +86.6\% & +5.2\% \\
    \midrule
    131K, 1 & \evalsecond{21,430} & 19,810 & 21,380 & \evalbest{23,030} & \evalsecond{21,014} & 20,699 & -3.4\% & -10.1\% \\
    131K, 4 & \evalsecond{22,059} & 20,673 & 21,619 & 33,089 & \evalsecond{33,662} & \evalbest{35,765} & +62.1\% & +6.2\% \\
    131K, 8 & \evalsecond{22,058} & 20,662 & 21,554 & 33,242 & \evalsecond{33,873} & \evalbest{36,030} & +63.3\% & +6.4\% \\
    131K, 16 & \evalsecond{22,052} & 20,670 & 21,677 & 33,322 & \evalsecond{33,984} & \evalbest{36,129} & +63.8\% & +6.3\% \\
    131K, 32 & \evalsecond{22,070} & 20,669 & 21,691 & 33,410 & \evalsecond{34,102} & \evalbest{35,878} & +62.6\% & +5.2\% \\
    \midrule
    262K, 1 & 18,772 & 17,785 & \evalsecond{18,938} & \evalsecond{21,479} & \evalbest{21,754} & 21,065 & +11.2\% & -3.2\% \\
    262K, 4 & 18,989 & 18,138 & \evalsecond{19,220} & 25,224 & \evalsecond{27,041} & \evalbest{27,924} & +45.3\% & +3.3\% \\
    262K, 8 & 18,987 & 18,133 & \evalsecond{19,230} & 25,271 & \evalsecond{27,110} & \evalbest{27,988} & +45.5\% & +3.2\% \\
    262K, 16 & 18,990 & 18,138 & \evalsecond{19,249} & 25,296 & \evalsecond{27,144} & \evalbest{27,971} & +45.3\% & +3.0\% \\
    262K, 32 & 18,995 & 18,118 & \evalsecond{19,254} & 25,321 & \evalsecond{27,179} & \evalbest{28,031} & +45.6\% & +3.1\% \\
    \bottomrule
  \end{tabular*}
\end{table}

\begin{table}[!htbp]
  \centering
  \caption{Queue-inclusive completion time for cold prefill, in seconds, with WavePP deltas against the lowest-latency TP8/EP8 and PP8 results in each cell. Negative deltas indicate lower completion time for WavePP.}
  \label{tab:cold-completion}
  \evaltablesetup
  \setlength{\tabcolsep}{0pt}
  \begin{tabular*}{\linewidth}{@{\extracolsep{\fill}}l
    r@{\hspace{1.5pt}}r
    r@{\hspace{1.5pt}}r
    r@{\hspace{1.5pt}}r
    @{\hspace{3pt}}|@{\hspace{3pt}}
    r@{\hspace{1.5pt}}r
    r@{\hspace{1.5pt}}r
    r@{\hspace{1.5pt}}r
    r@{\hspace{1.5pt}}r
    r@{\hspace{1.5pt}}r@{}}
    \toprule
    & \multicolumn{2}{c}{\shortstack{vLLM\\TP8/EP8}} &
      \multicolumn{2}{c}{\shortstack{SGLang\\TP8/EP8}} &
      \multicolumn{2}{c@{\hspace{3pt}}|@{\hspace{3pt}}}{\shortstack{TRT-LLM\\TP8/EP8}} &
      \multicolumn{2}{c}{\shortstack{vLLM\\PP8}} &
      \multicolumn{2}{c}{\shortstack{SGLang\\PP8}} &
      \multicolumn{2}{c}{WavePP} &
      \multicolumn{2}{c}{$\Delta_{\mathrm{TP}}$} &
      \multicolumn{2}{c}{$\Delta_{\mathrm{PP}}$} \\
    \cmidrule(lr){2-3}\cmidrule(lr){4-5}\cmidrule(lr){6-7}
    \cmidrule(lr){8-9}\cmidrule(lr){10-11}\cmidrule(lr){12-13}
    \cmidrule(lr){14-15}\cmidrule(lr){16-17}
    Context, $c$ & p50 & p95 & p50 & p95 & p50 & p95 & p50 & p95 & p50 & p95 & p50 & p95 &
      p50 & p95 & p50 & p95 \\
    \midrule
    65.5K, 1 & \evalbest{2.8} & \evalbest{2.8} & 3.2 & 3.2 & \evalsecond{2.9} & \evalsecond{2.9} & \evalsecond{3.2} & \evalsecond{3.2} & 3.9 & 3.9 & 3.0 & 3.0 & +7.1\% & +7.1\% & -6.2\% & -6.2\% \\
    65.5K, 4 & \evalsecond{10.8} & 11.5 & 11.8 & 11.8 & 11.1 & \evalsecond{11.1} & \evalbest{6.1} & \evalbest{6.1} & \evalsecond{6.3} & \evalsecond{6.3} & 6.8 & 7.1 & -37.0\% & -36.0\% & +11.5\% & +16.4\% \\
    65.5K, 8 & \evalsecond{21.6} & \evalsecond{22.3} & 23.5 & 23.8 & 22.3 & 22.5 & \evalsecond{12.2} & \evalbest{12.2} & 12.6 & 12.6 & \evalbest{11.8} & \evalsecond{12.4} & -45.4\% & -44.4\% & -3.3\% & +1.6\% \\
    65.5K, 16 & \evalsecond{43.9} & \evalsecond{44.0} & 47.1 & 47.3 & 44.5 & 44.5 & \evalsecond{24.4} & \evalsecond{24.4} & 25.1 & 25.1 & \evalbest{22.8} & \evalbest{23.2} & -48.1\% & -47.3\% & -6.6\% & -4.9\% \\
    65.5K, 32 & \evalsecond{87.3} & \evalsecond{87.9} & 94.4 & 94.4 & 89.0 & 89.1 & \evalsecond{48.8} & \evalsecond{48.8} & 50.1 & 50.2 & \evalbest{45.4} & \evalbest{45.9} & -48.0\% & -47.8\% & -7.0\% & -5.9\% \\
    \midrule
    131K, 1 & \evalsecond{6.1} & \evalsecond{6.2} & 6.6 & 6.6 & \evalsecond{6.1} & \evalsecond{6.2} & \evalbest{5.7} & \evalbest{5.7} & \evalsecond{6.2} & \evalsecond{6.2} & 6.3 & 6.4 & +3.3\% & +3.2\% & +10.5\% & +12.3\% \\
    131K, 4 & \evalsecond{24.0} & \evalsecond{24.1} & 25.3 & 25.3 & 24.2 & 24.2 & 15.6 & 15.6 & \evalsecond{15.3} & \evalsecond{15.3} & \evalbest{14.2} & \evalbest{14.8} & -40.8\% & -38.6\% & -7.2\% & -3.3\% \\
    131K, 8 & \evalsecond{47.4} & \evalsecond{48.1} & 50.7 & 50.9 & 48.4 & 49.8 & 31.2 & 31.3 & \evalsecond{30.6} & \evalsecond{30.6} & \evalbest{28.5} & \evalbest{29.1} & -39.9\% & -39.5\% & -6.9\% & -4.9\% \\
    131K, 16 & \evalsecond{94.8} & \evalsecond{95.5} & 101.3 & 101.6 & 96.7 & 96.8 & 62.4 & 62.5 & \evalsecond{61.1} & \evalsecond{61.1} & \evalbest{56.8} & \evalbest{57.6} & -40.1\% & -39.7\% & -7.0\% & -5.7\% \\
    131K, 32 & \evalsecond{189.6} & \evalsecond{190.3} & 202.9 & 202.9 & 193.2 & 193.5 & 124.9 & 124.9 & \evalsecond{122.2} & \evalsecond{122.2} & \evalbest{114.4} & \evalbest{116.3} & -39.7\% & -38.9\% & -6.4\% & -4.8\% \\
    \midrule
    262K, 1 & 14.0 & \evalsecond{14.0} & 14.7 & 14.7 & \evalsecond{13.8} & \evalsecond{14.0} & \evalsecond{12.2} & \evalsecond{12.2} & \evalbest{12.1} & \evalbest{12.1} & 12.4 & 12.8 & -10.1\% & -8.6\% & +2.5\% & +5.8\% \\
    262K, 4 & 55.2 & 55.2 & 57.8 & 57.8 & \evalsecond{54.6} & \evalsecond{54.6} & 41.3 & 41.3 & \evalsecond{38.5} & \evalsecond{38.5} & \evalbest{37.1} & \evalbest{37.7} & -32.1\% & -31.0\% & -3.6\% & -2.1\% \\
    262K, 8 & 110.4 & 110.5 & 115.6 & 115.8 & \evalsecond{109.0} & \evalsecond{109.1} & 82.6 & 82.7 & \evalsecond{77.0} & \evalsecond{77.0} & \evalbest{74.2} & \evalbest{74.8} & -31.9\% & -31.4\% & -3.6\% & -2.9\% \\
    262K, 16 & 220.8 & 220.9 & 231.1 & 231.4 & \evalsecond{217.7} & \evalsecond{218.2} & 165.3 & 165.3 & \evalsecond{153.9} & \evalsecond{153.9} & \evalbest{149.0} & \evalbest{149.7} & -31.6\% & -31.4\% & -3.2\% & -2.7\% \\
    262K, 32 & 441.0 & 441.7 & 462.5 & 462.7 & \evalsecond{435.4} & \evalsecond{435.9} & 330.5 & 330.6 & \evalsecond{307.8} & \evalsecond{307.9} & \evalbest{297.8} & \evalbest{298.9} & -31.6\% & -31.4\% & -3.2\% & -2.9\% \\
    \bottomrule
  \end{tabular*}
\end{table}

The higher throughput at $c\geq8$ is accompanied by lower completion time. WavePP's p50 is 31.6 to 49.0\% lower than TRT-LLM TP8/EP8 and lower than both PP baselines in every such setting. Its p95 is also lower than both PP baselines, except for cold 65.5K at $c=8$, where it is close to vLLM PP8 (1.6\% higher).

\subsection{Seeded Prefix Reuse}
\label{sec:eval-reuse}

We next consider requests with 90\% of the input cached. Each request then has much less prefill work left. This makes the dependence on concurrency more pronounced. Short suffixes provide few chunks per request, so several concurrent requests are needed to keep all stages busy. At $c=1$, all PP baselines have lower throughput than the TP baselines as most of the pipeline remains idle. Nevertheless, WavePP has the highest throughput among PP methods for both 131K and 262K requests, by about 31\% over vLLM PP8. Its throughput is still 63.3\% and 51.4\% below vLLM TP8/EP8, respectively. At $c=4$, WavePP's throughput remains below the TP baselines for 131K inputs, but for 262K inputs it is 15.3\% higher than TRT-LLM TP8/EP8 and 25.8\% higher than SGLang PP8. From $c=8$ onward, it has the highest throughput at both input lengths. Thus, the amount of work remaining after reuse is very important for pipeline occupancy.

For 131K reuse at $c\geq8$, WavePP improves throughput by 36.8 to 55.4\% over TRT-LLM TP8/EP8. At $c=32$, it reaches \num{289096} input tok/s with p50/p95 completion times of 13.2/17.0~s. Across all five concurrencies, including the low-concurrency settings, its geometric-mean throughput is 4.1\% below TRT-LLM TP8/EP8 but 11.6\% and 24.9\% above SGLang PP8 and vLLM PP8.

For 262K reuse, WavePP improves geometric-mean throughput by 11.1\% over TRT-LLM TP8/EP8, 22.2\% over SGLang PP8, and 45.3\% over vLLM PP8. At $c=32$, it serves \num{218743} input tok/s with p50/p95 completion times of 35.8/39.7~s, compared with \num{184753} input tok/s and 44.1/44.9~s for SGLang PP8.

\begin{figure}[!htbp]
  \centering
  \includegraphics[width=\linewidth]{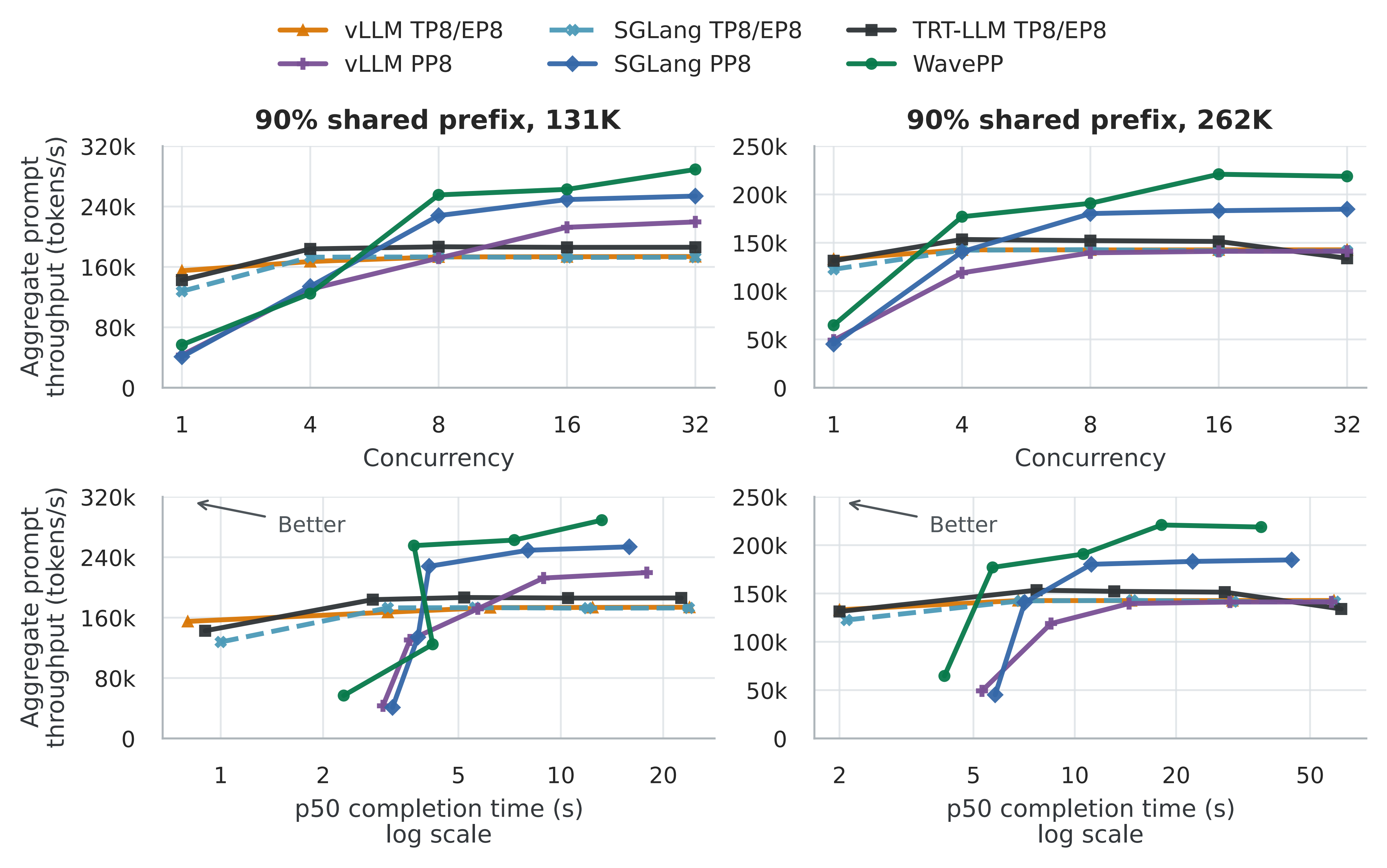}
  \caption{Seeded reuse with a 90\% shared prefix. The top row shows throughput by concurrency. The bottom row shows throughput p50 operating points. The arrow in each operating-point panel points toward higher throughput and lower completion time.}
  \label{fig:reuse-performance}
\end{figure}

\begin{table}[!htbp]
  \centering
  \caption{Seeded-reuse aggregate prompt throughput (input tok/s). Positive deltas indicate higher throughput for WavePP.}
  \label{tab:reuse-throughput}
  \evaltablesetup
  \setlength{\tabcolsep}{2pt}
  \begin{tabular*}{\linewidth}{@{\extracolsep{\fill}}lrrr@{\hspace{3pt}}|@{\hspace{3pt}}rrrrr@{}}
    \toprule
    Context, $c$ & \shortstack{vLLM\\TP8/EP8} & \shortstack{SGLang\\TP8/EP8} &
      \shortstack{TRT-LLM\\TP8/EP8} & \shortstack{vLLM\\PP8} &
      \shortstack{SGLang\\PP8} & WavePP &
      $\Delta_{\mathrm{TP}}$ &
      $\Delta_{\mathrm{PP}}$ \\
    \midrule
    131K, 1 & \evalbest{155,002} & 127,746 & \evalsecond{142,607} & \evalsecond{43,197} & 41,022 & 56,810 & -63.3\% & +31.5\% \\
    131K, 4 & 167,165 & \evalsecond{172,924} & \evalbest{183,856} & \evalsecond{130,420} & 134,161 & 124,682 & -32.2\% & -7.1\% \\
    131K, 8 & 173,271 & 173,266 & \evalsecond{186,690} & 171,714 & \evalsecond{228,012} & \evalbest{255,406} & +36.8\% & +12.0\% \\
    131K, 16 & 173,510 & 172,581 & \evalsecond{185,931} & 212,477 & \evalsecond{249,227} & \evalbest{262,740} & +41.3\% & +5.4\% \\
    131K, 32 & 173,780 & 172,847 & \evalsecond{186,064} & 219,649 & \evalsecond{253,844} & \evalbest{289,096} & +55.4\% & +13.9\% \\
    \midrule
    262K, 1 & \evalbest{133,140} & 122,545 & \evalsecond{131,382} & \evalsecond{49,238} & 45,127 & 64,674 & -51.4\% & +31.3\% \\
    262K, 4 & 142,686 & 142,359 & \evalsecond{153,418} & 118,899 & \evalsecond{140,713} & \evalbest{176,960} & +15.3\% & +25.8\% \\
    262K, 8 & 142,617 & 142,854 & \evalsecond{152,211} & 139,579 & \evalsecond{180,199} & \evalbest{190,842} & +25.4\% & +5.9\% \\
    262K, 16 & 142,661 & 141,581 & \evalsecond{151,376} & 141,134 & \evalsecond{183,182} & \evalbest{220,899} & +45.9\% & +20.6\% \\
    262K, 32 & \evalsecond{142,715} & 141,497 & 133,965 & 141,248 & \evalsecond{184,753} & \evalbest{218,743} & +53.3\% & +18.4\% \\
    \bottomrule
  \end{tabular*}
\end{table}

\begin{table}[!htbp]
  \centering
  \caption{Queue-inclusive completion time for seeded reuse, in seconds, with WavePP deltas against the lowest-latency TP8/EP8 and PP8 results in each cell. Negative deltas indicate lower completion time for WavePP.}
  \label{tab:reuse-completion}
  \evaltablesetup
  \setlength{\tabcolsep}{0pt}
  \begin{tabular*}{\linewidth}{@{\extracolsep{\fill}}l
    r@{\hspace{1.5pt}}r
    r@{\hspace{1.5pt}}r
    r@{\hspace{1.5pt}}r
    @{\hspace{3pt}}|@{\hspace{3pt}}
    r@{\hspace{1.5pt}}r
    r@{\hspace{1.5pt}}r
    r@{\hspace{1.5pt}}r
    r@{\hspace{1.5pt}}r
    r@{\hspace{1.5pt}}r@{}}
    \toprule
    & \multicolumn{2}{c}{\shortstack{vLLM\\TP8/EP8}} &
      \multicolumn{2}{c}{\shortstack{SGLang\\TP8/EP8}} &
      \multicolumn{2}{c@{\hspace{3pt}}|@{\hspace{3pt}}}{\shortstack{TRT-LLM\\TP8/EP8}} &
      \multicolumn{2}{c}{\shortstack{vLLM\\PP8}} &
      \multicolumn{2}{c}{\shortstack{SGLang\\PP8}} &
      \multicolumn{2}{c}{WavePP} &
      \multicolumn{2}{c}{$\Delta_{\mathrm{TP}}$} &
      \multicolumn{2}{c}{$\Delta_{\mathrm{PP}}$} \\
    \cmidrule(lr){2-3}\cmidrule(lr){4-5}\cmidrule(lr){6-7}
    \cmidrule(lr){8-9}\cmidrule(lr){10-11}\cmidrule(lr){12-13}
    \cmidrule(lr){14-15}\cmidrule(lr){16-17}
    Context, $c$ & p50 & p95 & p50 & p95 & p50 & p95 & p50 & p95 & p50 & p95 & p50 & p95 &
      p50 & p95 & p50 & p95 \\
    \midrule
    131K, 1 & \evalbest{0.8} & \evalbest{0.9} & 1.0 & \evalsecond{1.0} & \evalsecond{0.9} & \evalbest{0.9} & \evalsecond{3.0} & \evalsecond{3.0} & 3.2 & 3.2 & 2.3 & 2.3 & +187.5\% & +155.6\% & -23.3\% & -23.3\% \\
    131K, 4 & \evalsecond{3.1} & \evalsecond{3.7} & \evalsecond{3.1} & 3.8 & \evalbest{2.8} & \evalbest{3.0} & 3.6 & \evalsecond{4.6} & \evalsecond{3.8} & 4.7 & 4.2 & 4.5 & +50.0\% & +50.0\% & +16.7\% & -2.2\% \\
    131K, 8 & 6.2 & \evalsecond{6.3} & 5.6 & 7.0 & \evalsecond{5.2} & 6.9 & 5.7 & 7.9 & \evalsecond{4.1} & \evalsecond{6.2} & \evalbest{3.7} & \evalbest{5.5} & -28.8\% & -12.7\% & -9.8\% & -11.3\% \\
    131K, 16 & 12.4 & 12.5 & 12.0 & 12.7 & \evalsecond{10.5} & \evalsecond{12.3} & 8.9 & 12.4 & \evalsecond{8.0} & \evalbest{9.3} & \evalbest{7.3} & \evalsecond{9.6} & -30.5\% & -22.0\% & -8.8\% & +3.2\% \\
    131K, 32 & 23.9 & 24.8 & 23.8 & 24.7 & \evalsecond{22.6} & \evalsecond{22.8} & 17.9 & 20.3 & \evalsecond{15.9} & \evalbest{16.6} & \evalbest{13.2} & \evalsecond{17.0} & -41.6\% & -25.4\% & -17.0\% & +2.4\% \\
    \midrule
    262K, 1 & \evalbest{2.0} & \evalbest{2.0} & \evalsecond{2.1} & \evalsecond{2.1} & \evalbest{2.0} & \evalbest{2.0} & \evalsecond{5.3} & \evalsecond{5.3} & 5.8 & 5.8 & 4.1 & 4.1 & +105.0\% & +105.0\% & -22.6\% & -22.6\% \\
    262K, 4 & \evalsecond{6.8} & \evalsecond{8.0} & 6.9 & 8.1 & 7.7 & 8.5 & 8.5 & 10.2 & \evalsecond{7.1} & \evalsecond{8.4} & \evalbest{5.7} & \evalbest{6.2} & -16.2\% & -22.5\% & -19.7\% & -26.2\% \\
    262K, 8 & 14.7 & \evalsecond{14.7} & 14.8 & 15.0 & \evalsecond{13.1} & 15.2 & 14.5 & 16.4 & \evalsecond{11.2} & \evalbest{12.4} & \evalbest{10.6} & \evalsecond{12.7} & -19.1\% & -13.6\% & -5.4\% & +2.4\% \\
    262K, 16 & 29.4 & 29.5 & 29.5 & 30.9 & \evalsecond{27.9} & \evalsecond{28.4} & 28.9 & 29.0 & \evalsecond{22.4} & \evalsecond{22.4} & \evalbest{18.1} & \evalbest{19.3} & -35.1\% & -32.0\% & -19.2\% & -13.8\% \\
    262K, 32 & \evalsecond{58.9} & \evalsecond{58.9} & 59.3 & 60.6 & 61.9 & 66.0 & 57.9 & 58.0 & \evalsecond{44.1} & \evalsecond{44.9} & \evalbest{35.8} & \evalbest{39.7} & -39.2\% & -32.6\% & -18.8\% & -11.6\% \\
    \bottomrule
  \end{tabular*}
\end{table}

At $c\geq8$, WavePP reduces p50 completion time by 19.1--42.2\% relative to TRT-LLM TP8/EP8 and by 5.4--19.2\% relative to the PP baseline with the lower p50. Its p95 is lower than both PP baselines in three of the six settings and 2.4--3.2\% higher than the lower PP baseline p95 in the other three. WavePP therefore improves throughput and median completion time more consistently than p95 completion time.

\subsection{Mixed Traffic and Completion Latency}
\label{sec:eval-mixed}

Mixed traffic places short uncached suffixes alongside long cold requests. WavePP has the highest throughput at all three concurrencies. It improves geometric-mean throughput by 46.1\% over TRT-LLM TP8/EP8, 56.8\% over SGLang TP8/EP8, and 50.0\% over vLLM TP8/EP8. Relative to the PP baselines, WavePP improves geometric-mean throughput by 5.0\% over SGLang and 14.6\% over vLLM. At $c=32$, WavePP serves \num{61190} input tok/s with p50/p95 completion times of 83.8/109.7~s.

\begin{figure}[!htbp]
  \centering
  \includegraphics[width=\linewidth]{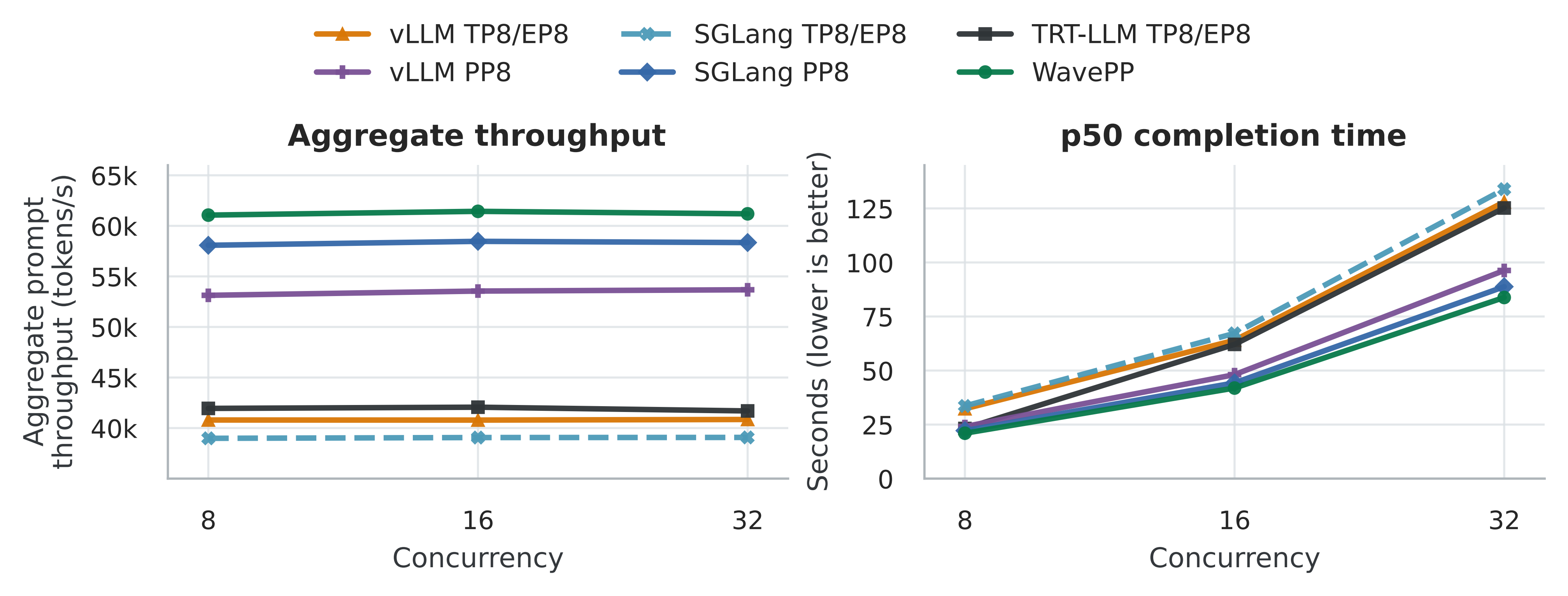}
  \caption{Mixed 75/25 warm-131K/cold-262K traffic. The left panel reports aggregate throughput and the right reports queue-inclusive p50 completion time.}
  \label{fig:mixed-performance}
\end{figure}

\begin{table}[!htbp]
  \centering
  \caption{Mixed-workload aggregate prompt throughput (input tok/s). Positive deltas indicate higher throughput for WavePP.}
  \label{tab:mixed-throughput}
  \evaltablesetup
  \setlength{\tabcolsep}{2pt}
  \begin{tabular*}{\linewidth}{@{\extracolsep{\fill}}rrrr@{\hspace{3pt}}|@{\hspace{3pt}}rrrrr@{}}
    \toprule
    $c$ & \shortstack{vLLM\\TP8/EP8} & \shortstack{SGLang\\TP8/EP8} &
      \shortstack{TRT-LLM\\TP8/EP8} & \shortstack{vLLM\\PP8} &
      \shortstack{SGLang\\PP8} & WavePP &
      $\Delta_{\mathrm{TP}}$ &
      $\Delta_{\mathrm{PP}}$ \\
    \midrule
    8 & 40,797 & 38,986 & \evalsecond{41,945} & 53,129 & \evalsecond{58,079} & \evalbest{61,063} & +45.6\% & +5.1\% \\
    16 & 40,790 & 39,062 & \evalsecond{42,064} & 53,551 & \evalsecond{58,468} & \evalbest{61,436} & +46.1\% & +5.1\% \\
    32 & 40,849 & 39,074 & \evalsecond{41,695} & 53,678 & \evalsecond{58,348} & \evalbest{61,190} & +46.8\% & +4.9\% \\
    \bottomrule
  \end{tabular*}
\end{table}

\begin{table}[!htbp]
  \centering
  \caption{Queue-inclusive completion time for mixed traffic, in seconds, with WavePP deltas against the lowest-latency TP8/EP8 and PP8 results in each cell. Negative deltas indicate lower completion time for WavePP.}
  \label{tab:mixed-completion}
  \evaltablesetup
  \setlength{\tabcolsep}{0pt}
  \begin{tabular*}{\linewidth}{@{\extracolsep{\fill}}r
    r@{\hspace{1.5pt}}r
    r@{\hspace{1.5pt}}r
    r@{\hspace{1.5pt}}r
    @{\hspace{3pt}}|@{\hspace{3pt}}
    r@{\hspace{1.5pt}}r
    r@{\hspace{1.5pt}}r
    r@{\hspace{1.5pt}}r
    r@{\hspace{1.5pt}}r
    r@{\hspace{1.5pt}}r@{}}
    \toprule
    & \multicolumn{2}{c}{\shortstack{vLLM\\TP8/EP8}} &
      \multicolumn{2}{c}{\shortstack{SGLang\\TP8/EP8}} &
      \multicolumn{2}{c@{\hspace{3pt}}|@{\hspace{3pt}}}{\shortstack{TRT-LLM\\TP8/EP8}} &
      \multicolumn{2}{c}{\shortstack{vLLM\\PP8}} &
      \multicolumn{2}{c}{\shortstack{SGLang\\PP8}} &
      \multicolumn{2}{c}{WavePP} &
      \multicolumn{2}{c}{$\Delta_{\mathrm{TP}}$} &
      \multicolumn{2}{c}{$\Delta_{\mathrm{PP}}$} \\
    \cmidrule(lr){2-3}\cmidrule(lr){4-5}\cmidrule(lr){6-7}
    \cmidrule(lr){8-9}\cmidrule(lr){10-11}\cmidrule(lr){12-13}
    \cmidrule(lr){14-15}\cmidrule(lr){16-17}
    $c$ & p50 & p95 & p50 & p95 & p50 & p95 & p50 & p95 & p50 & p95 & p50 & p95 &
      p50 & p95 & p50 & p95 \\
    \midrule
    8 & 32.3 & \evalsecond{45.2} & 33.6 & 47.5 & \evalsecond{23.2} & 46.6 & 24.1 & \evalsecond{26.4} & \evalsecond{22.2} & 31.3 & \evalbest{21.0} & \evalbest{25.6} & -9.5\% & -43.4\% & -5.4\% & -3.0\% \\
    16 & 64.0 & 77.6 & 67.2 & 67.9 & \evalsecond{62.1} & \evalsecond{66.5} & 48.1 & \evalsecond{50.7} & \evalsecond{44.3} & \evalbest{44.7} & \evalbest{41.9} & 51.1 & -32.5\% & -23.2\% & -5.4\% & +14.3\% \\
    32 & 128.0 & \evalsecond{141.6} & 133.9 & 148.7 & \evalsecond{125.2} & 142.4 & 96.3 & 115.9 & \evalsecond{88.8} & \evalbest{98.1} & \evalbest{83.8} & \evalsecond{109.7} & -33.1\% & -22.5\% & -5.6\% & +11.8\% \\
    \bottomrule
  \end{tabular*}
\end{table}

WavePP reduces p50 completion time by 9.5--33.1\% and p95 by 22.5--43.4\% relative to the lowest corresponding TP baseline values. Its p50 is also 5.4--5.6\% lower than the lowest PP baseline p50. Relative to the lowest PP baseline p95, WavePP's p95 is 3.0\% lower at $c=8$, but 14.3\% and 11.8\% higher at $c=16$ and $c=32$. In this mixed workload, WavePP improves throughput and reduces median completion time, while p95 completion time is higher at the two larger concurrencies. 

\subsection{Prefix Retention}
\label{sec:eval-retention}

To measure how well reusable state survives as new requests arrive, we compare WavePP with TRT-LLM TP8/EP8 at five arrival rates between 0.50 and 1.00 requests/s. A good hit reuses at least 80\% of the expected prefix. WavePP achieves a good-hit rate of 88--100\%, compared with 73--83\% for the tensor-parallel baseline. Its good-hit rate is higher at every tested rate.

\begin{figure}[!htbp]
  \centering
  \includegraphics[width=0.74\linewidth]{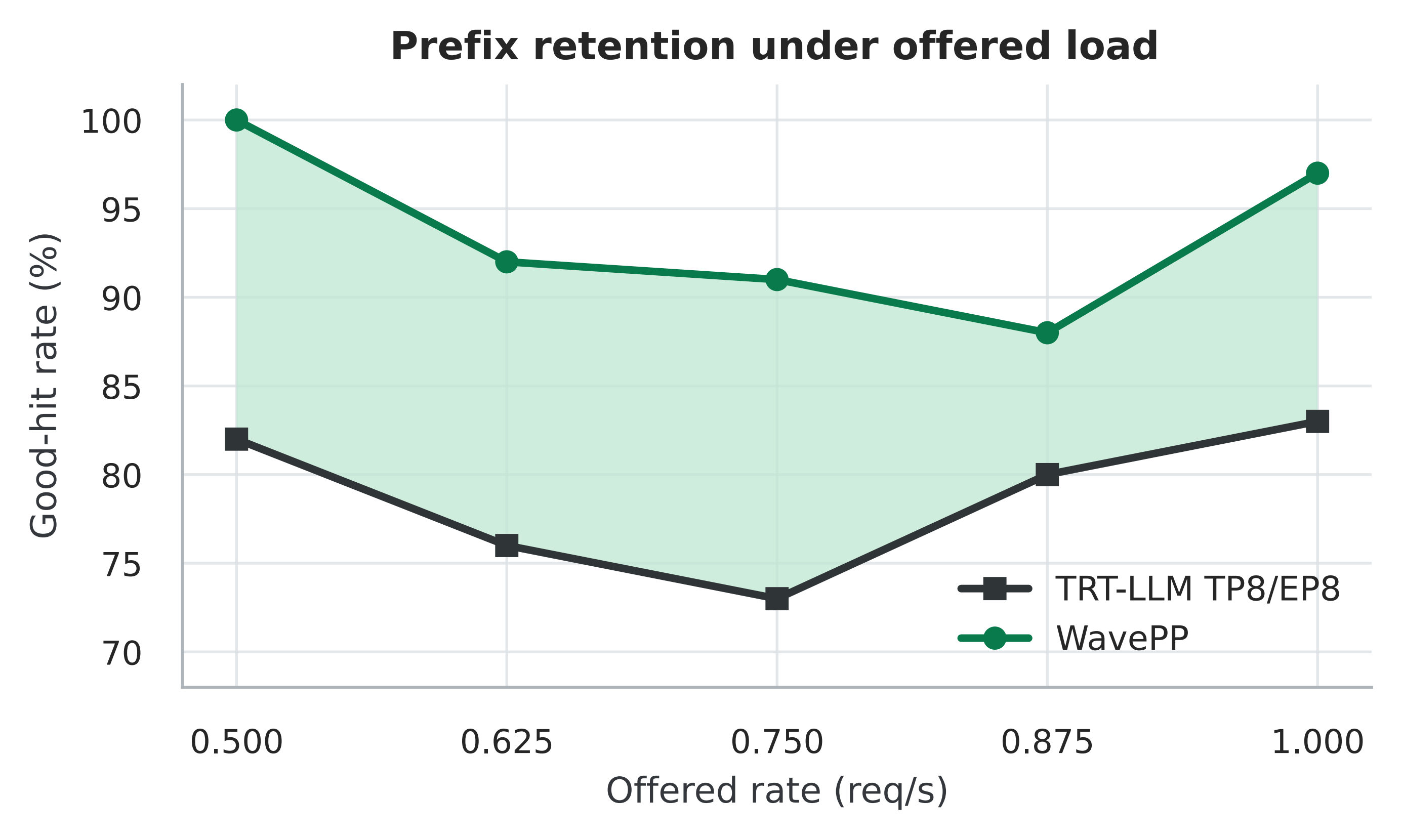}
  \caption{Good-hit rate in the prefix-retention sweep.  Both systems receive
  identical requests at each offered-load setting.}
  \label{fig:prefix-retention}
\end{figure}

\begin{table}[!htbp]
  \centering
  \caption{Prefix-retention good-hit rate.  The five offered-load settings
  span 0.50--1.00 requests/s; positive differences indicate a higher good-hit rate for WavePP.}
  \label{tab:prefix-retention}
  \evaltablesetup
  \begin{tabular*}{\linewidth}{@{\extracolsep{\fill}}rrrr@{}}
    \toprule
    Offered rate (req/s) & TRT-LLM TP8/EP8 & WavePP &
      \shortstack{Difference (pp)\\($\uparrow$ better)} \\
    \midrule
    0.500 & \evalsecond{82\%} & \evalbest{100\%} & +18 \\
    0.625 & \evalsecond{76\%} & \evalbest{92\%} & +16 \\
    0.750 & \evalsecond{73\%} & \evalbest{91\%} & +18 \\
    0.875 & \evalsecond{80\%} & \evalbest{88\%} & +8 \\
    1.000 & \evalsecond{83\%} & \evalbest{97\%} & +14 \\
    \bottomrule
  \end{tabular*}
\end{table}

\subsection{Ablation Study}
\label{sec:eval-ablation}

We evaluate the admission and scheduling mechanisms on one eight-GPU B300 node using the same code with different mechanisms enabled. Each configuration ran the six workloads in one server boot. Table~\ref{tab:ablation-rungs} lists the five configurations. Starting from our naive PP8 re-implementation, we add wave admission and transport, then leases and capacity escrows, and finally MPU. The MPU phase includes planner-driven packing, adaptive wave sizing, and the progress deadline, so its results measure these scheduling changes together.

\begin{figure}[!htbp]
  \centering
  \includegraphics[width=\linewidth]{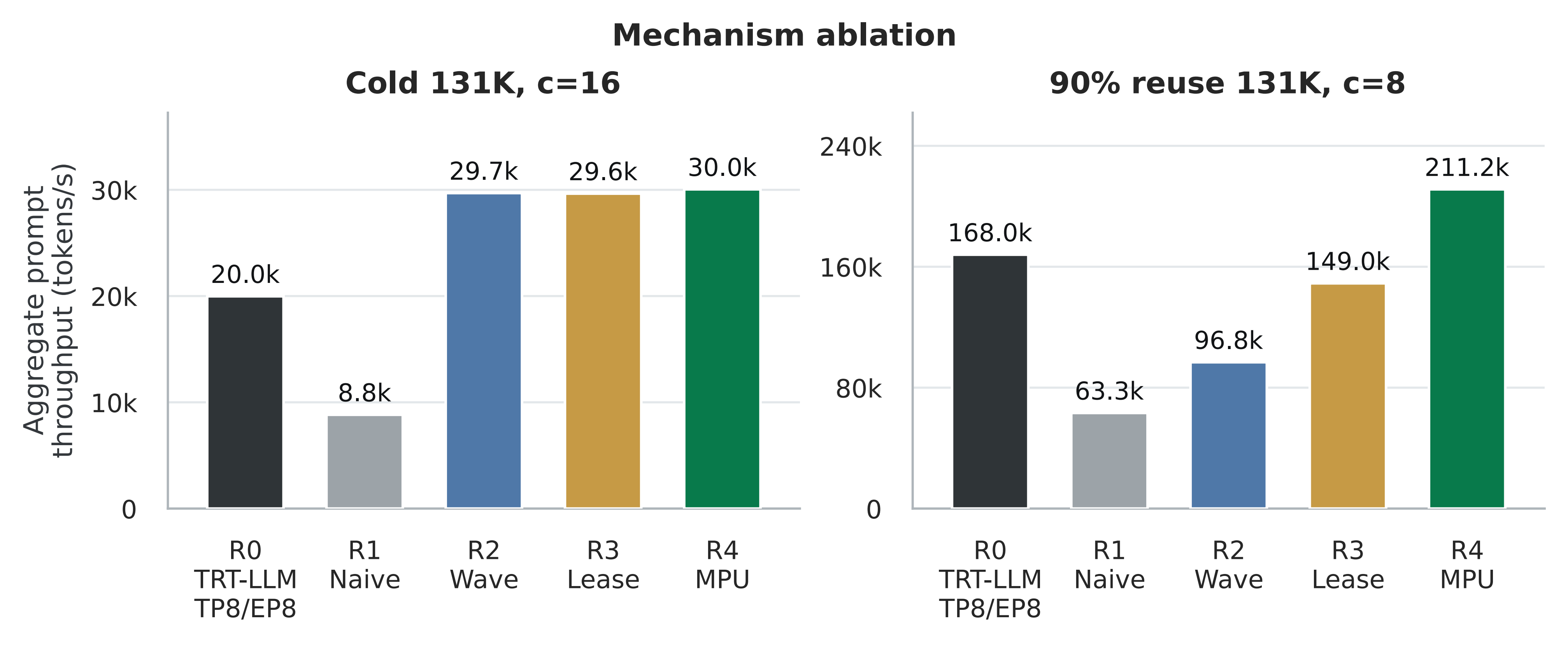}
  \caption{Cumulative B300 ablation in two representative cells.  Bars report aggregate throughput for the tensor-parallel reference and each cumulative configuration. R1 is the naive re-implementation. R4 combines planner-driven packing, adaptive wave sizing, and a progress deadline as MPU.}
  \label{fig:mechanism-ablation}
\end{figure}

The TP8/EP8 reference uses a 32,768-token budget and a batch limit of 32, which we found to be the best setting we tested. All PP configurations use a 16,384-token budget and a batch limit of 16. The comparison with TP therefore includes these configuration differences, while the successive PP comparisons keep them fixed. Figure~\ref{fig:mechanism-ablation} shows cold 131K at $c=16$ and reuse 131K at $c=8$.

\begin{table}[!htbp]
  \centering
  \caption{Cumulative ablation configurations.}
  \label{tab:ablation-rungs}
  \evaltablesetup
  \begin{tabularx}{\linewidth}{@{}lp{0.28\linewidth}X@{}}
    \toprule
    Rung & Configuration & Added mechanism \\
    \midrule
    R0 & TRT-LLM TP8/EP8 & Tensor-parallel reference \\
    R1 & Naive re-implementation & Lockstep request preparation \\
    R2 & + wave admission/transport & Independent context-wave execution \\
    R3 & + lease admission & Exact reuse and capacity ownership \\
    R4 & + MPU & Planner-driven packing, adaptive wave sizing, and progress deadline \\
    \bottomrule
  \end{tabularx}
\end{table}

On cold 131K at $c=16$, wave admission and transport increase throughput from \num{8826} to \num{29692} input tok/s, a 236.4\% increase. Later changes have little effect on throughput. MPU reduces p50 completion time from 69.69 to 43.03~s, but mean completion time changes only from 63.18 to 62.33~s and throughput rises by just 1.4\%. 

Admission and scheduling both matter more for cached requests. On reuse 131K at $c=8$, adding leases and capacity escrows increases throughput from \num{96796} to \num{149021} input tok/s, a 54.0\% increase. MPU further increases throughput to \num{211214}, an additional 41.7\%, while reducing p50 from 6.51 to 4.20~s. This is because, without these mechanisms, stages can wait for reusable state causing gaps between GPU work. On the other hand, adaptive wave sizing spreads the limited suffix work across more waves so that more stages can execute concurrently. Both mechanisms matter here because requests have little computation left after reuse.

On reuse 262K at $c=32$, the combined MPU phase reduces throughput from \num{175305} to \num{158238} input tok/s, a 9.7\% decrease, although throughput remains 24.5\% above TP8/EP8. Thus, in some circumstances, planner-driven packing can limit performance gains as compared to the optimal packing strategy (as we see a static packing performs better) for a specific workload, though the adaptibility is still beneficial for mixed and ever-changing workloads.

\FloatBarrier

\section{Limitations and Further Directions}
\label{sec:limitations}

\paragraph{Fully dynamic chunking}
MPU adjusts the token budget in discrete steps based on the amount of admitted work. It does not estimate the execution time of each chunk. Yet equally sized chunks can have different costs because they attend to different prefix lengths or encounter different stage bottlenecks. A more flexible policy could choose chunk sizes from estimated execution times and observed stage progress. The challenge is to improve pipeline balance and completion latency while keeping planning inexpensive enough to stay ahead of execution. We further observed that adding MPU decreases performance in certain experiments. Improving the MPU policy to be more adaptive could mitigate these regressions.

\paragraph{Other pipeline topologies}
The current evaluation uses TP1$\times$PP4 and TP1$\times$PP8. Configurations such as TP2$\times$PP4 and PP16 would help establish how admission overhead and overlap change with stage count, tensor-parallel collectives, and inter-node bandwidth, and how other topologies respond to WavePP. Uneven layer placement is another useful case, since keeping requests ready cannot by itself remove a persistent imbalance in stage execution time.

\paragraph{Scheduling and cache policies}
WavePP currently packs requests in FCFS order and reserves capacity before scheduling their chunks. Further work could explore alternative scheduling strategies that consider request priorities, deadlines, and the dynamic state of the cache to optimize both latency and throughput.

\paragraph{Multiple replicas}
Combining multiple replicas with cache-aware routing across prefill workers could improve tail latency. Evaluating WavePP and the baselines in this setting under realistic traffic patterns would also test the benefits of a larger cache size for a long-running deployment. This could also include producing detailed SLO curves to better understand the latency behavior under different load conditions.

\FloatBarrier
\section{Conclusion}

Pipeline parallelism requires more than keeping the GPUs busy with forward passes. New requests must also find reusable state and obtain enough cache capacity while earlier requests are still running. This becomes a distributed operation when stages retain, offload, and evict their caches independently. Performing this preparation on the execution path can delay the next wave even when the GPUs have work available.

We presented WavePP, which moves reuse discovery and admission ahead of execution. Hints reduce the cache-tree walks needed to find a common reusable prefix. Leases protect the chosen prefix, while capacity escrows reserve space for the suffix without immediately evicting reusable blocks. Each stage then materializes the request before executing its scheduled chunks. These mechanisms allow cache management to proceed independently across stages while the pipeline follows a common schedule.

With the same kernels and PP4 topology, WavePP improves TensorRT-LLM's prefill throughput in 37 of 40 GLM~5.2 and MiniMax M2.7 settings. At concurrency 128, throughput for short-suffix requests under cache pressure increases by factors of $2.91$ and $2.02$. WavePP also achieves the highest measured throughput in all 18 Kimi K3 settings at concurrency eight or higher. The ablations show that admission and transport improve cold prefill, while coordinated cache admission and scheduling provide additional gains under reuse. Together, these results show the value of preparing requests early and keeping enough work available to sustain pipeline execution.

\section{Acknowledgements}

The author thanks Joyjit Daw and Philip Howes for their tremendous support and encouragement throughout the project, and Joyjit for the helpful discussions.

The author used AI tools to improve the clarity and presentation of the manuscript and to assist with generation of figures and tables. The research ideas, methodology, and experimental results are the author's own work. The author reviewed the manuscript and takes full responsibility for its content.

\bibliographystyle{waveppnat}
\bibliography{wavepp_cited}

\end{document}